\documentclass[fleqn,usenatbib]{mnras}

\usepackage{newtxtext,newtxmath}

\usepackage[T1]{fontenc}
\usepackage{ae,aecompl}

\usepackage{graphicx}	
\usepackage{amsmath}	
\usepackage{amssymb}	
\pdfoutput=1 

\title[3D plasma motions]{Analysis of 3D plasma motions in a chromospheric jet formed due to magnetic reconnection}

\author[J. J. Gonz\'alez-Avil\'es et al.]{
J. J. Gonz\'alez-Avil\'es,$^{1}$\thanks{E-mail:jjgonzalez@igeofisica.unam.mx (JJGA)}
        F. S. Guzm\'an,$^{2}$
        V. Fedun,$^{3}$
        G.Verth,$^{4}$ 
        R. Sharma,$^{5}$
        S. Shelyag $^{6}$ 
        \newauthor
        and S. Regnier$^{6}$
        \\
        $^{1}$Instituto de Geof\'isica, Unidad Michoac\'an, Universidad Nacional Aut\'onoma de M\'exico, Antigua Carretera a P\'atzcuaro, 8710 Morelia, Michoac\'an, M\'{e}xico \\
        $^2$Laboratorio de Inteligencia Artificial y Superc\'omputo, Instituto de F\'{\i}sica y Matem\'{a}ticas, Universidad \\
        Michoacana de San Nicol\'as de Hidalgo. Edificio C3, Cd. Universitaria, 5840 Morelia, Michoac\'{a}n, M\'{e}xico \\
        $^3$ Department of Automatic Control and Systems Engineering, The University of Sheffield, Mappin Street, Sheffield, S1 3JD, UK \\
        $^4$ School of Mathematics and Statistics, The University of Sheffield, Hicks Building, Hounsfield Road, Sheffield, S3 7RH, UK \\
        $^5$ Space Research Group - Space Weather, Departamento de F\'isica y Matem\'aticas, Universidad de Alcal\'a, Calle el Escorial, 19-21, 28805 Alcal\'a de Henares, Spain \\
        $^6$ Department of Mathematics, Physics and Electrical Engineering, Northumbria University, Ellison Place, Newcastle upon Tyne, NE1 8ST, UK
        }

\date{Accepted XXX. Received YYY; in original form ZZZ}

\pubyear{2018}

\begin{document}
\label{firstpage}
\pagerange{\pageref{firstpage}--\pageref{lastpage}}
\maketitle

\begin{abstract}
Within the framework of resistive MHD, implementing the C7 equilibrium atmosphere model and a 3D potential magnetic field realistic configuration, we simulate the formation of a plasm
jet with the morphology, upward velocity up to 130 km/s and timescale formation between 60 and 90 s after beginning of simulation, similar to those expected for Type II spicules. Initial
results of this simulation were published in Paper \citep[e.g.,][]{Gonzalez-Aviles_et_al_2018} and present paper is devoted to the analysis of transverse displacements and rotational type
motion of the jet. Our results suggest that 3D magnetic reconnection may be responsible for the formation of the jet in Paper \citep{Gonzalez-Aviles_et_al_2018}. In this paper, by calculating times series of
the velocity components $v_x$ and $v_y$ in different points near to the jet for various heights we find transverse oscillations in agreement with spicule observations. We also obtain a time-distance plot of
the temperature in a cross-cut at the plane $x=$0.1 Mm and find significant transverse displacements of the jet. By analyzing temperature isosurfaces of $10^{4}$ K with the distribution of $v_x$, we find
that if the line-of-sight (LOS) is approximately perpendicular to the jet axis then there is both motion towards and away from the observer across the width of the jet. This red-blue shift pattern of the jet is
caused by rotational motion, initially clockwise and anti-clockwise afterwards, which could be interpreted as torsional motion. From a nearly vertical perspective of the jet the LOS velocity component shows
a central blue-shift region surrounded by red-shifted plasma. 
\end{abstract}




\section{Introduction}
\label{sec:introduction}

In the solar atmosphere, jet-like structures, defined as an impulsive evolution of collimated bright or dark structure are observed in a wide range of environments. In particular, the upper
chromosphere is full with spicules, thin jets of chromospheric plasma that reach heights of 10,000 km or move above the photosphere. Although spicules were described since 1878 by
Secchi \citet{Secchi_1878}, understanding their physical nature has been a whole area of research \citep{Beckers_1968,Sterling_2000}. There are two types of spicules, the first type of
spicules are so-called Type I, which reach maximum heights of 4-8 Mm, maximum ascending velocities of 15-40 km s$^{-1}$, have a lifetime of 3-6.5 minutes \citep{Pereira_et_al_2012},
and show up and downward motions \citep{Beckers_1968,Suematsu_et_al_1995}. These Type I spicules are probably the counterpart of the dynamic fibrils. They follow a parabolic
(ballistic) path in space and time. In general the dynamics of these spicules is produced by mangneto-acoustic shock waves passing or wave-driving through the chromosphere
\citep{Shibata_et_al_1982,De_Pontieu_et_al_2004,Hansteen_et_al_2006,Martinez-Sykora_et_al_2009,Matsumoto&Shibata_2010,Scullion_et_al_2011}.The second type of spicules are called Type II,
which reach maximum heights of 3-9 Mm (longer in coronal holes) and have lifetimes of 50-150 s, shorter than that of Type I spicules \citep{De_Pontieu_et_al_2007a,Pereira_et_al_2012}. These Type II
spicules show apparent upward motion with speeds of order 30-110 km s$^{-1}$. At the end of their life they usually exhibit rapid fading in chromospheric lines \citep{De_Pontieu_et_al_2007b}. It has been
suggested from observations that Type II spicules are continuously accelerated while being heated to at least transition region temperatures \citep{De_Pontieu_et_al_2009,De_Pontieu_et_al_2011}. Other
observations indicate that some Type II spicules also show an increase or a more complex velocity dependence with height \citep{Sekse_et_al_2012}. 

Also, Type II spicules show other motions in addition to radial outflow. In the Ca II H line they are seen to sway transversely with amplitudes of order 10-20 km s$^{-1}$ and periods of 100-500 s
\citep{De_Pontieu_et_al_2007b,Tomczyk_et_al_2007,Zaqarashvili&Erdelyi_2009,McIntosh_et_al_2011,Sharma_2017}, suggesting generation of upward, downward and standing Alfv\'en waves
\citep{Okamoto&De_Pontieu_2011,Tavabi_et_al_2015}, the generation of MHD kink mode waves or Alfv\'en waves due to magnetic reconnection
\citep{Nishizuka_et_al_2008,He_et_al_2009,McLaughlin_et_al_2012,Kuridze_et_al_2012} or due to magnetic tension and ambipolar difussion \citep{Martinez-Sykora_et_al_2017}. For instance,
\citet{Suematsu_et_al_2008} suggest that some spicules show multi-thread structure as result of possible rotation. Another possible motion of  Type II spicules is the torsional one as suggested by
\citet{Beckers_1972} and \citet{Kayshap_et_al_2018}, and established using high-resolution spectroscopy at the limb \citep{De_Pontieu_et_al_2012}. According to the latter, Type II spicules show torsional
motions rotational speeds of 25-30 km s$^{-1}$. In addition, the continuation of this kind of motion in the transition region and coronal lines suggest that they may help driving the solar wind
\citep{McIntosh_et_al_2011}.

There are other types of motion less well established, for instance \citet{Curdt&Tian_2011} and \citet{Curdt_et_al_2012} suggest that the spinning motion of Type II spicules can explain the tilts of ultraviolet
lines in the so-called explosive events producing larger-scale macro spicules. These spectral-line tilts were observed at the limb and also attributed to spicule rotation \citep{Beckers_1972}. At smaller
scales, evidence of rotating motions has been deduced for the chromospheric/transition region jet events \citep{Liu_et_al_2009,Liu_et_al_2011}. In addition, \citet{Tian_et_al_2014} using the IRIS
instrument found transverse motions as well as line broadening attributed to the existence of twist and torsional Alfv\'en waves. At the photospheric level, there is evidence that a fraction of spicules present
twisting motions \citep{Sterling_et_al_2010a,Sterling_et_al_2010b,De_Pontieu_et_al_2012}. Beyond the resolution of imaging instruments, the spectrum of explosive events can also be interpreted as
arising from the fast rotation of magnetic structures \citep{Curdt&Tian_2011,Curdt_et_al_2012}. Apart from the small scale jets, Doppler images have shown that several coronal jet events present strong
rotational motion, diagnosed with blue-red shift  observed  on opposite sides of each jet \citep{Dere_et_al_1989,Pike&Mason_1998,Cheung_et_al_2015}. 

In this paper, we show that the jet with characteristics of a Type II spicule, obtained in the numerical simulations presented in \cite{Gonzalez-Aviles_et_al_2018} shows transverse displacements and
rotational type motion initially clockwise and anti-clockwise afterwards.

The summary of the model and numerical methods are described in Section \ref{sec:model_numerical_methods}. Section \ref{sec:Results} describes the analysis of the plasma motions in the jet. In
Section \ref{sec:conclusions}, we present our final comments and conclusions.


\section{Summary of the Model and Numerical Methods}
\label{sec:model_numerical_methods}

The details of the numerical methods can be found in \citet{Gonzalez-Aviles_et_al_2018} and a brief summary is the following. We solve the resistive 3D MHD equations including the
constant gravity field at the Sun's surface. We integrate the Extended Generalized Lagrange Multiplier (EGLM) resistive MHD \citep{Jiang_et_al_2012} using High Resolution Shock
Capturing methods with an adaptive choice of Flux formula between HLLC and HLLE, combined with MINMOD and MC limiters. 

For the initial magnetic field, we use a 3D potential (current-free) configuration extrapolated from a simulated quiet-Sun photospheric field, obtained from a large-scale, high-resolution, self-consistent
simulation of solar magnetoconvection in a bipolar photospheric region with the MURaM code \citep{Vogler_et_al_2005,Shelyag_et_al_2012}. The computational box has a size of 480$\times$480
$\times$400 pixels, with a spatial resolution of 25 km in all directions.  

\begin{figure}
\centering
\includegraphics[width=6.5cm]{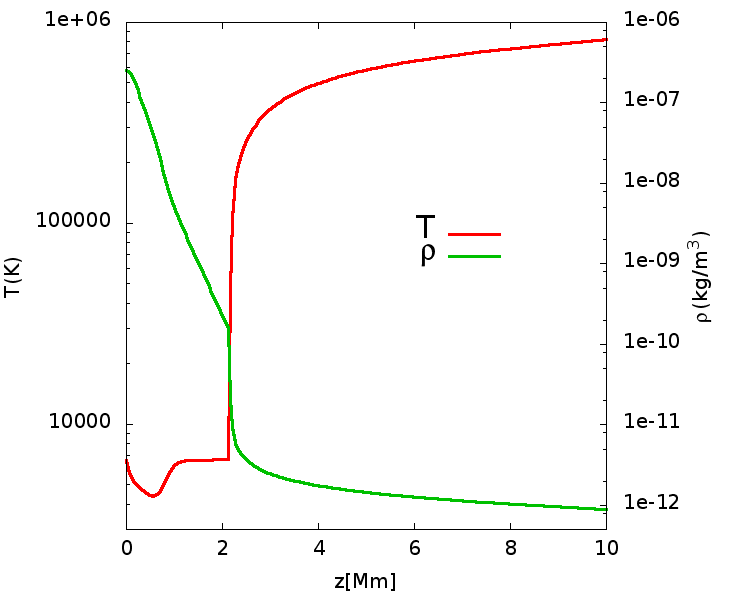}
\caption{Temperature and mass density as a function of height $z$ for the C7 equilibrium solar atmosphere model.}
\label{fig:atmosphere}
\end{figure} 

In order to model the atmosphere we choose the numerical domain to cover part of the interconnected solar photosphere, chromosphere and corona (see Fig.~\ref{fig:atmosphere}).
For this the atmosphere is initially assumed to be in hydrostatic equilibrium. The temperature field is considered to obey the semi-empirical C7 model of the chromosphere transition
region \citep{Avrett&Loeser2008} and is distributed consistently with observed line intensities and  profiles from the SUMER atlas of the extreme ultraviolet spectrum
\citep{Curdt_et_al_1999}. The photosphere is extended to the solar corona as described by \citet{Fontela_et_al_1990} and \citet{Griffiths_et_al_1999}. The temperature $T(z)$ and
mass density $\rho(z)$ are functions of  height $z$ and are shown in Fig. \ref{fig:atmosphere}, where the transition region shows its characteristic steep gradient.

Once the magnetic field and atmosphere model are set in the computational domain (240$\times$240$\times$400 grid cells with a resolution of 25 km in each direction), the plasma evolves due to the
inclusion of resistivity according to the EGLM equations. For our analysis, we focus on a 3D numerical box with unigrid discretization of size $x\in[0,6]$, $y\in[0,6]$, $z\in[0,10]$ Mm. In Section
\ref{sec:Results}, we analyze the transverse and rotational motions in the jet and their observational signatures. 


\section{Plasma motions in the jet}
\label{sec:Results}


\subsection{Transverse motions} 

\begin{figure*}
\centering
\includegraphics[scale=0.4]{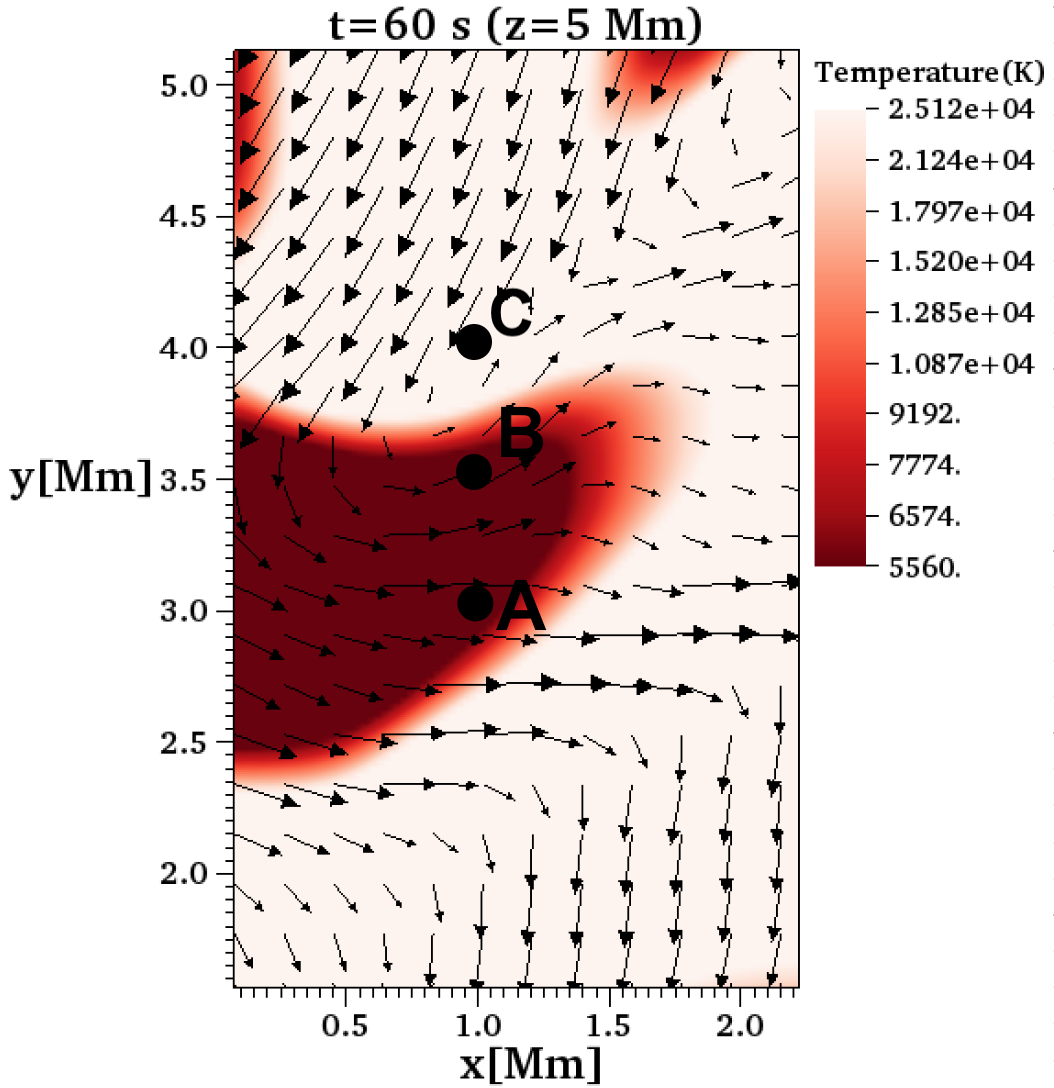}\\
\includegraphics[width=5cm]{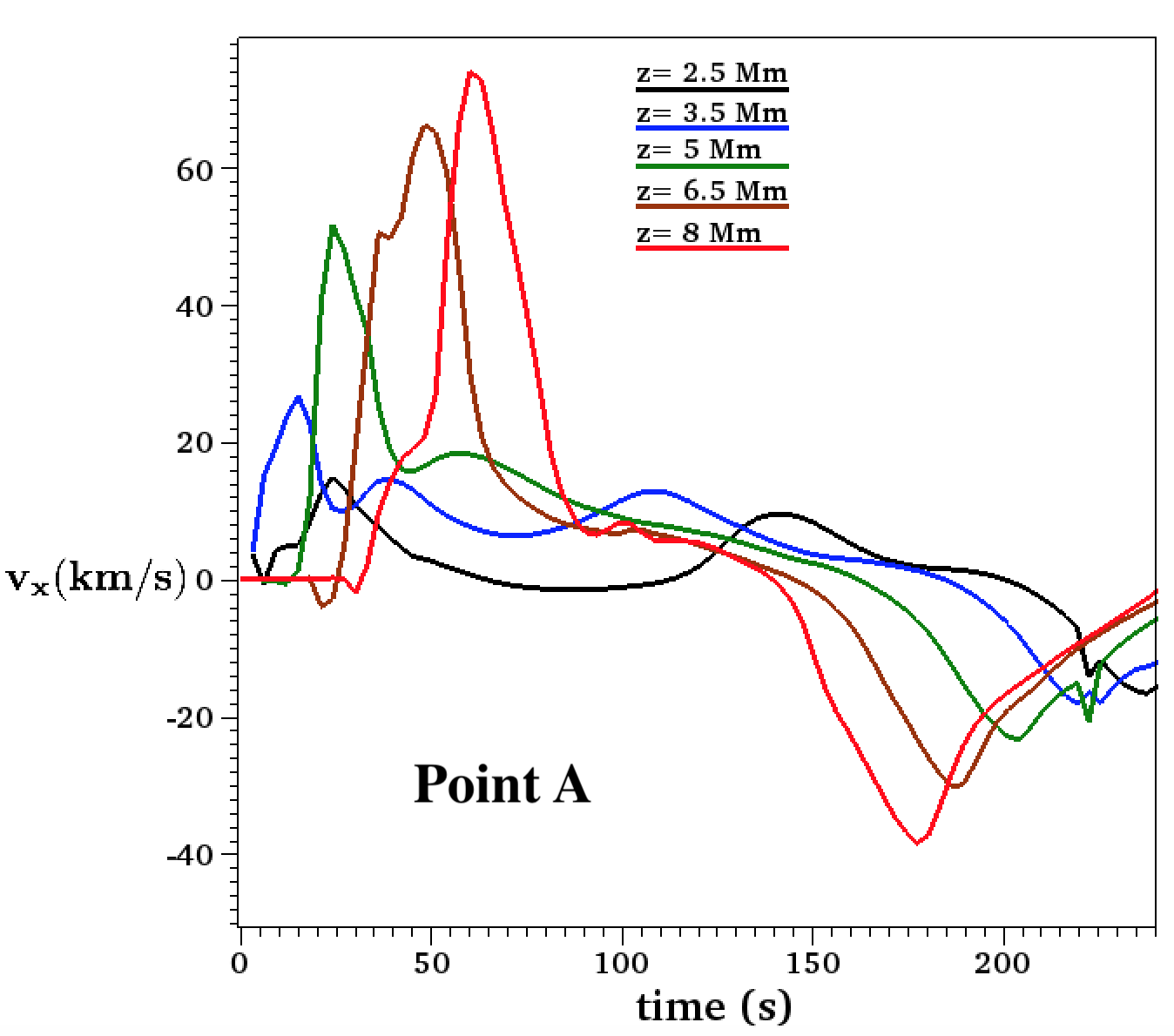}
\includegraphics[width=5cm]{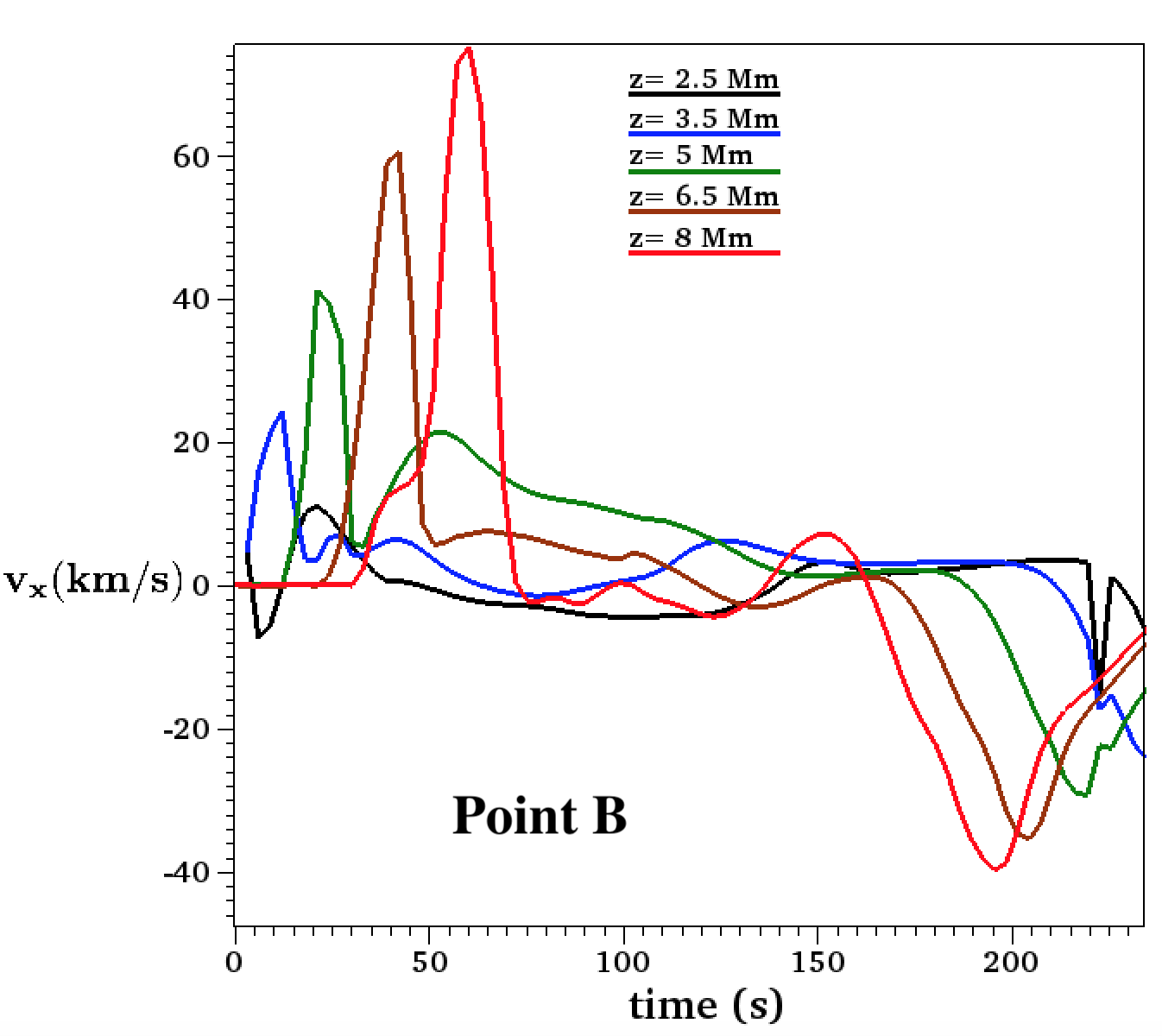}
\includegraphics[width=5cm]{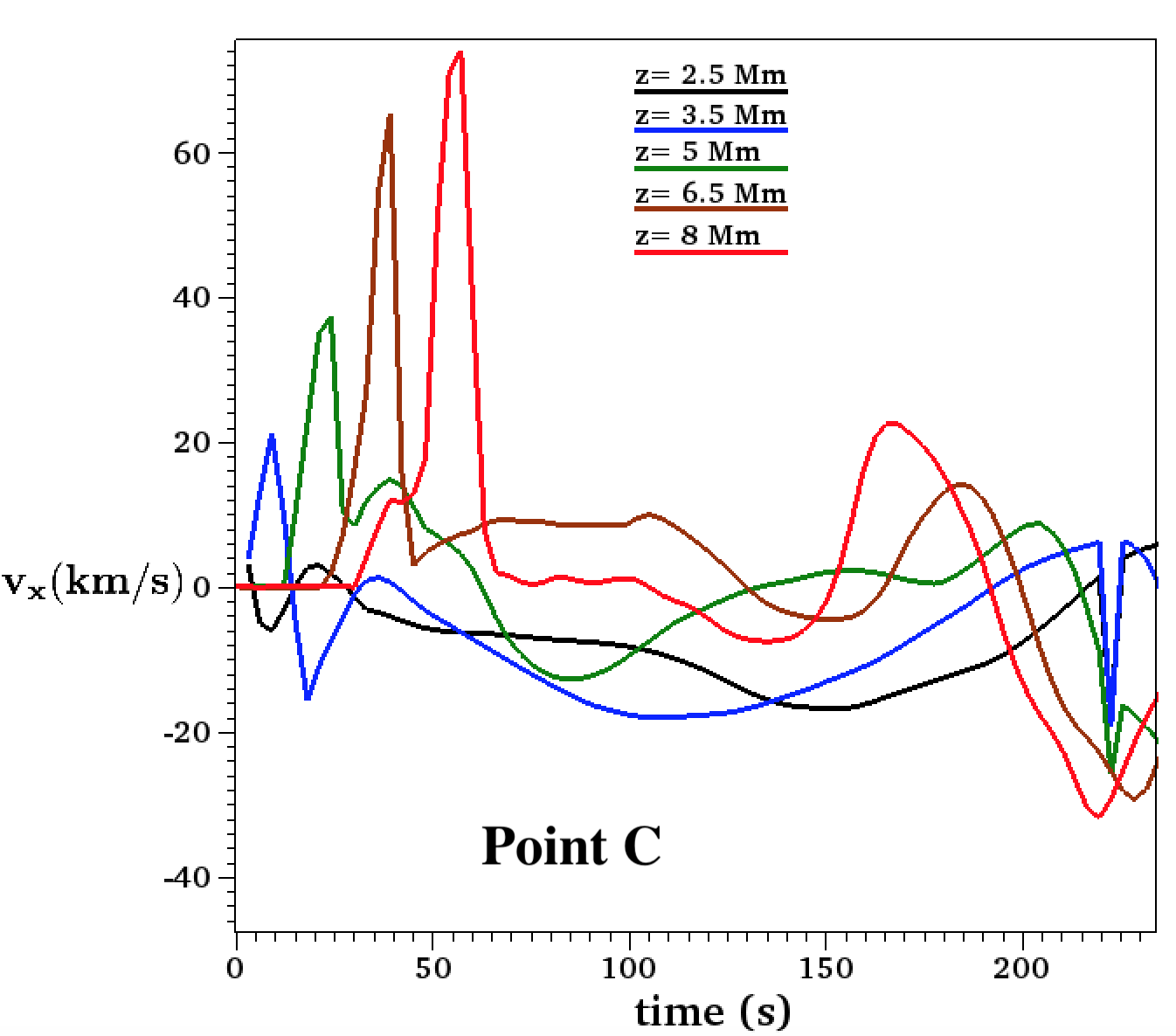}
\includegraphics[width=5cm]{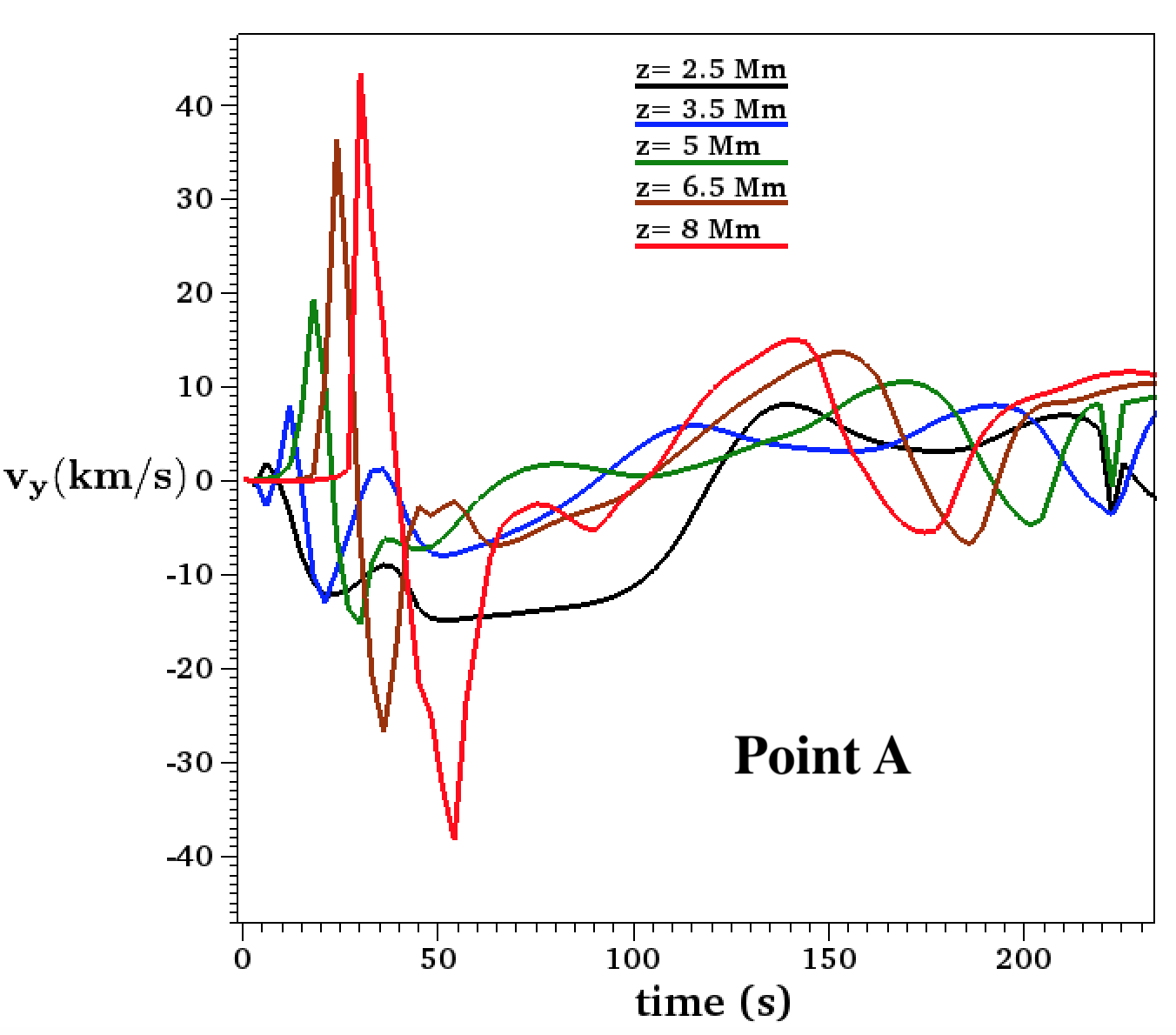}
\includegraphics[width=5cm]{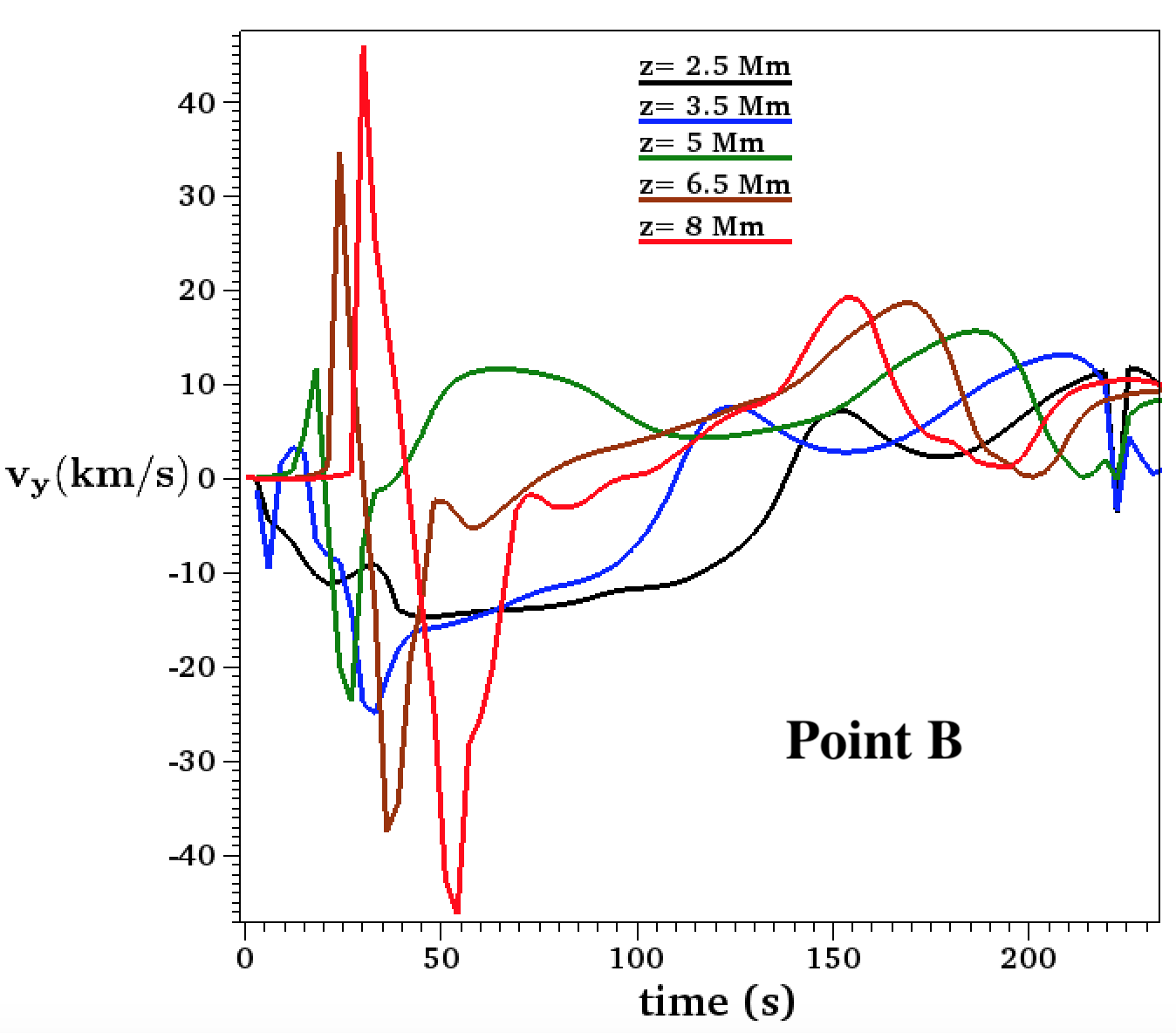}
\includegraphics[width=5cm]{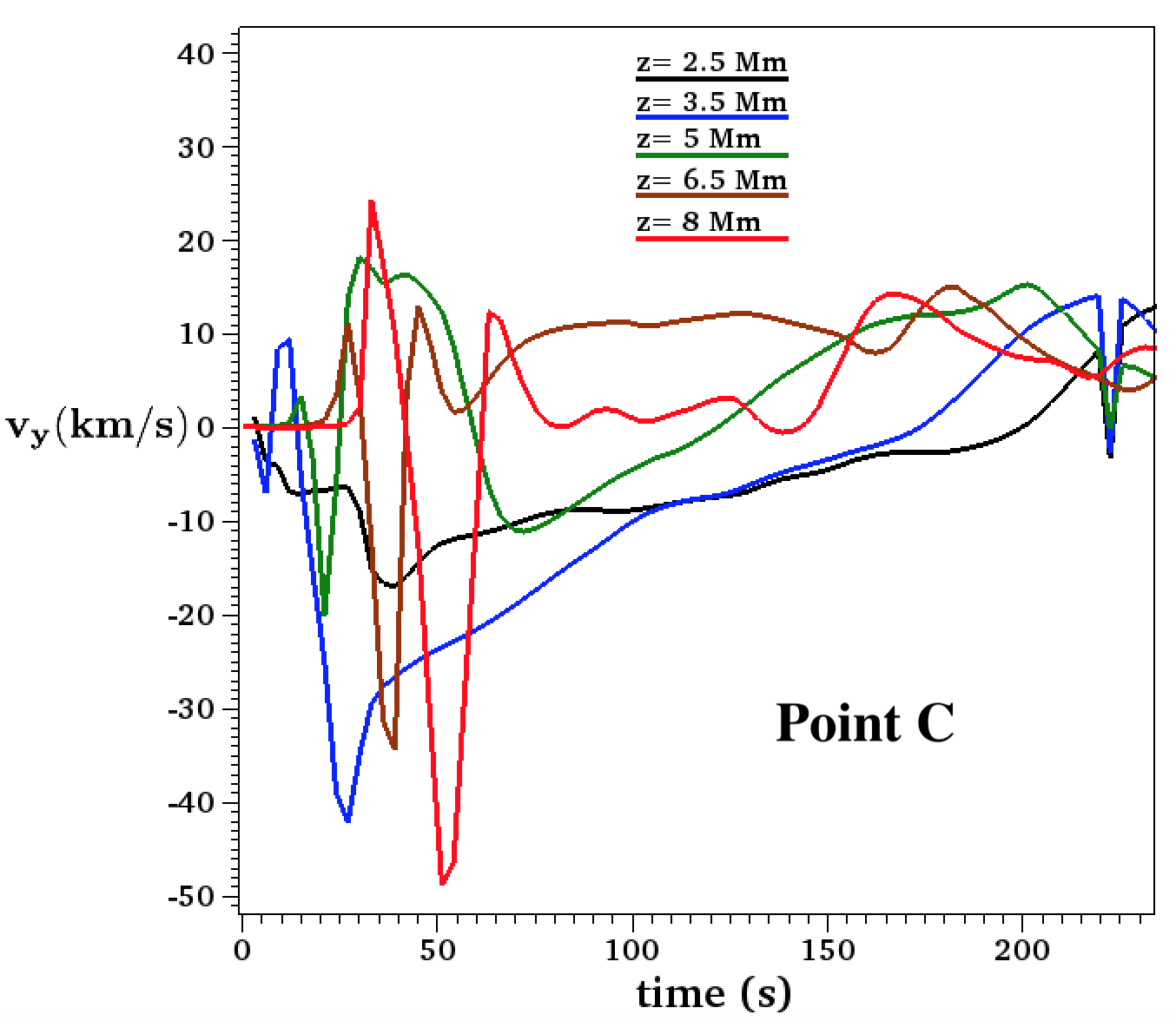}
\caption{\label{velocity_components_over_time} In the top we show the region where $v_x$ and $v_y$ are measured. The color labels the temperature in the plane $z=5$ Mm at time $t=60$ s, where the
structure of spicule and the circulation of the vector velocity field is clearly seen. In the middle and bottom panels we show the time series of $v_x$ and $v_y$ in km s$^{-1}$ of the volume elements at the
points A, B and C measured at various planes of constant height.}
\end{figure*}

A property to look into is the transverse displacement of the jet investigate if it is actually oscillating in a kink-like manner. We measure the velocity components $v_x$ and $v_y$ in time at three different
points near and within the spicule, points A ($x=$1, $y=3$, $z$) Mm, points B ($x=$1, $y=$3.5, $z$) Mm and points C ($x=1$, $y=4$, $z$) Mm, for various values of $z$. Near these points  the vector
velocity field rotates as is illustrated at the top of Fig. \ref{velocity_components_over_time}. We measure the value of the velocity components at heights  $z=$2.5, 3.5, 5, 6.5 and 8 Mm. The results
displayed in Fig. \ref{velocity_components_over_time} tell us about the motion along the $x$ and $y$ directions. For example the horizontal component of velocity $v_x$ at the various heights in point A
shows transverse displacements with high amplitudes at the top of jet and small  at the bottom. We can also see a change of sign, which indicates a transverse oscillation of at least one period. The
behavior of $v_x$ at points B and C is similar to that at point A. In the case of the $v_y$ component at point A, we can see strong motions at the top, in particular there is a clear change of sing between 0
and 100 s, then we can identify oscillatory behavior at all heights, which is also clear at points B and C. By comparing  the behavior of $v_x$ and $v_y$ we can conclude that the jet shows rotational
motion, i.e. velocity components are out of phase, which will be reinforced by the following analysis. 

\begin{figure*}
\centering
\includegraphics[width=5.5cm]{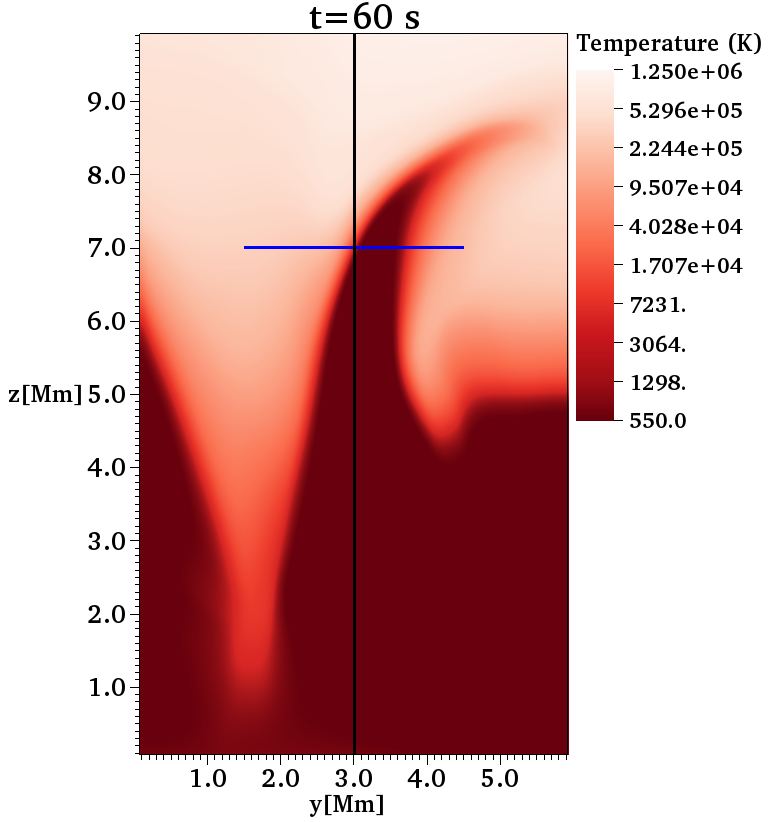}
\includegraphics[scale=0.36]{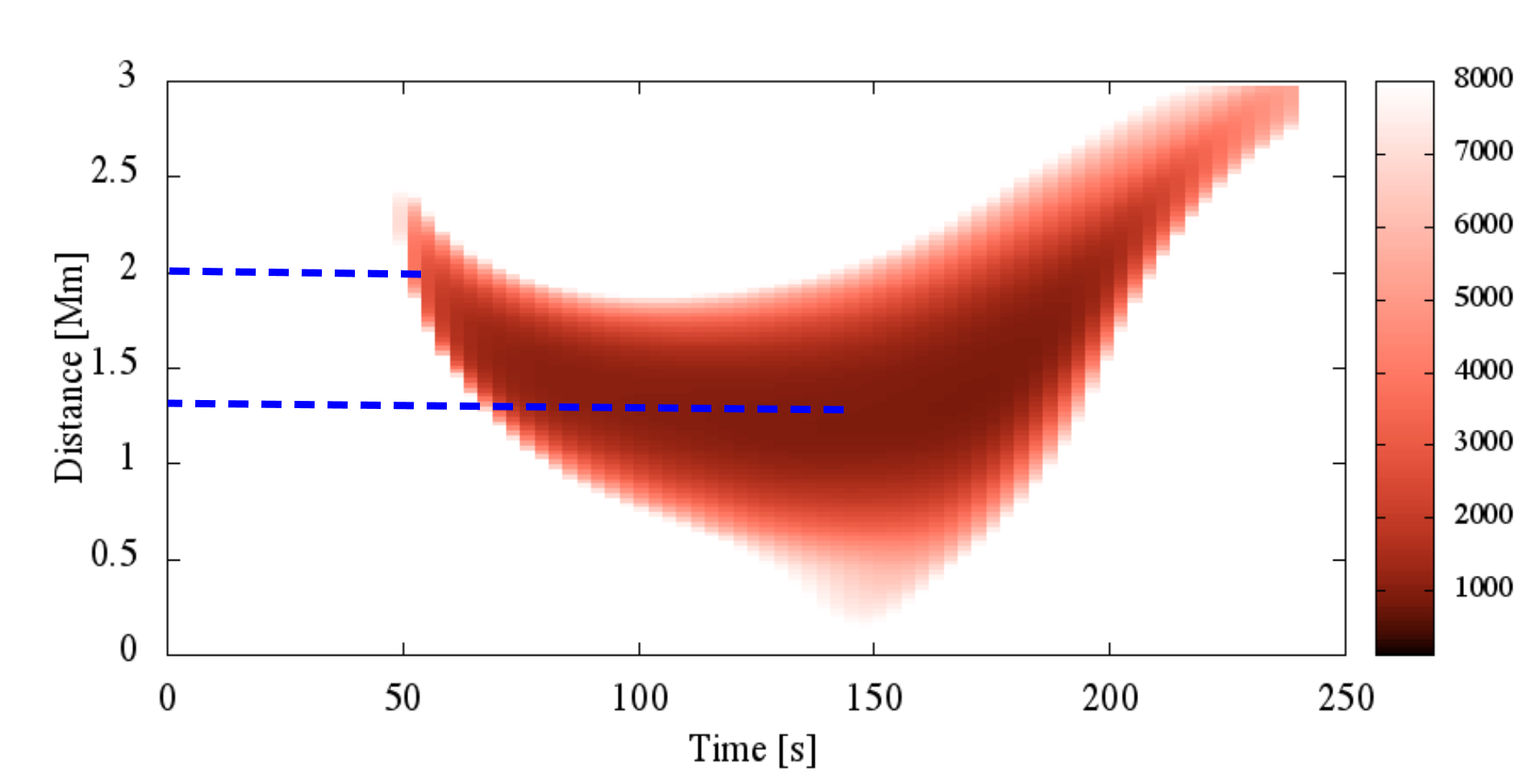}
\caption{\label{Time_distance_plots}(Left) Snapshot of the logarithm of temperature (K) at time $t=60$ s, vertical line in black at $y=3$ Mm and horizontal line in blue from $y=$ 1.5 Mm to $y=$ 4.5 Mm at
$z=7$ Mm. (Right) The time-distance plot of logarithm of temperature (K) and dashed lines to estimate the average transverse speed.}
\end{figure*}

Aside from analyzing the transverse motion at individual points in space, we can also identify the bulk transverse displacement of the jet in a horizontal cross-cut across the jet shown in a logarithm of
temperature on the left of Fig. \ref{Time_distance_plots}. We measure this bulk transverse motion at a height of 7 Mm along a horizontal slice of length 3 Mm (blue line) centered at the mid-point  of the
domain in the y-direction (black line) as shown on the left of Fig. \ref{Time_distance_plots}. A time-distance plot of the logarithm of temperature along this slice as a function of time is shown on the right of
Fig. \ref{Time_distance_plots}. From the time-distance plot we can see that from time $t=50$ s the jet starts moving to the left until about time $t=150$ s, jet starts moving to the right until it is displaced a
horizontal distance of 3 Mm. This shows that simulated jet actually has a significant transverse motion during its lifetime. This phenomena is also observed widely in spicule observations,
\citep[e.g.,][]{De_Pontieu_et_al_2007b}. To estimate the average speed of the transverse displacements, we indicate the center of the jet at the the times of maximum and minimum displacement up to 150 s with horizontal dashed blue lines on the right of Fig. \ref{Time_distance_plots}. The distance between the two lines is about 0.7 Mm (700 km) and the time between them is about 100 s, therefore the average speed is about 7 km s$^{-1}$.  

\subsection{Rotational motions}

\begin{figure*}
\centering
\includegraphics[width=5.5cm]{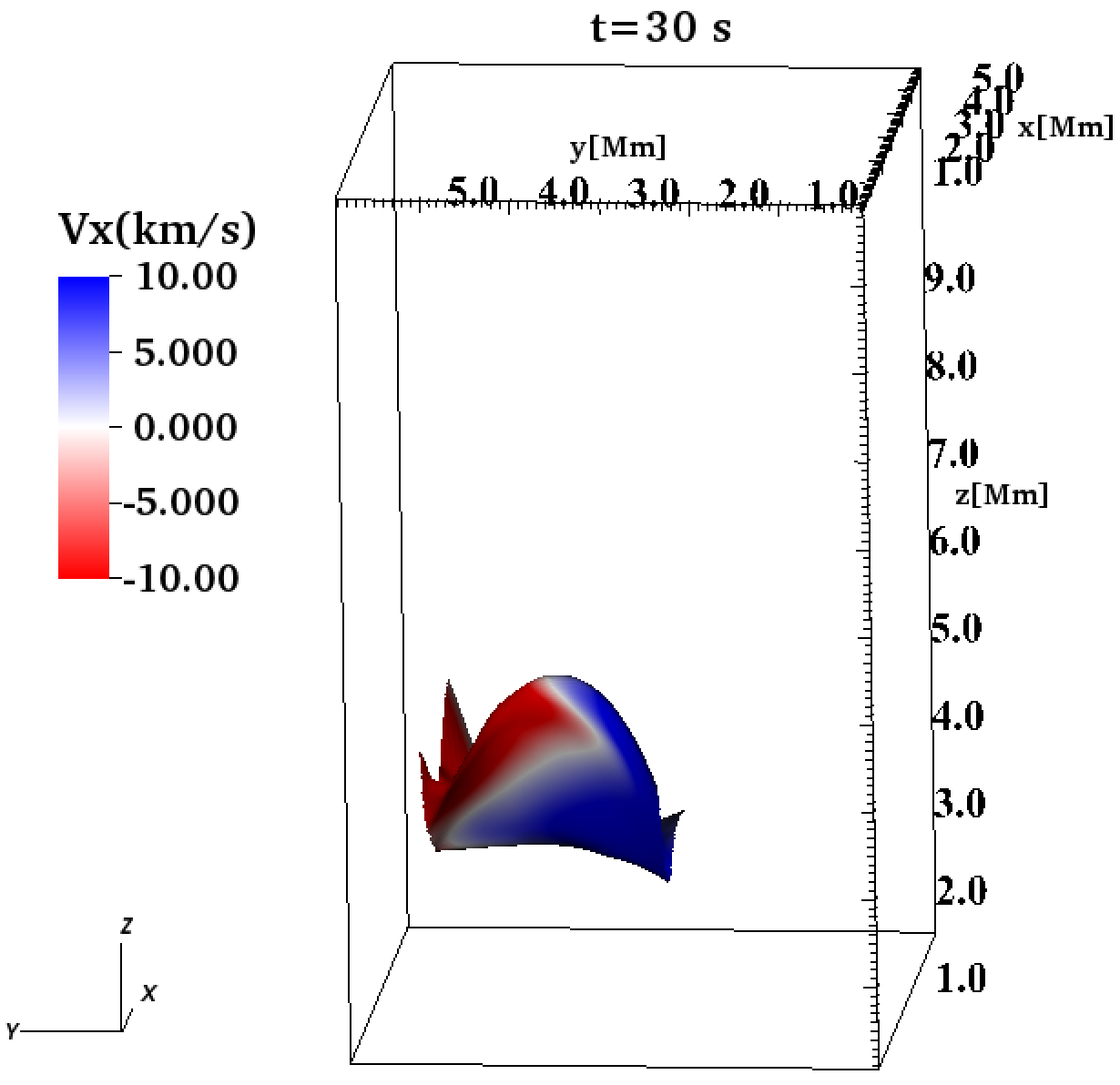}
\includegraphics[width=5.5cm]{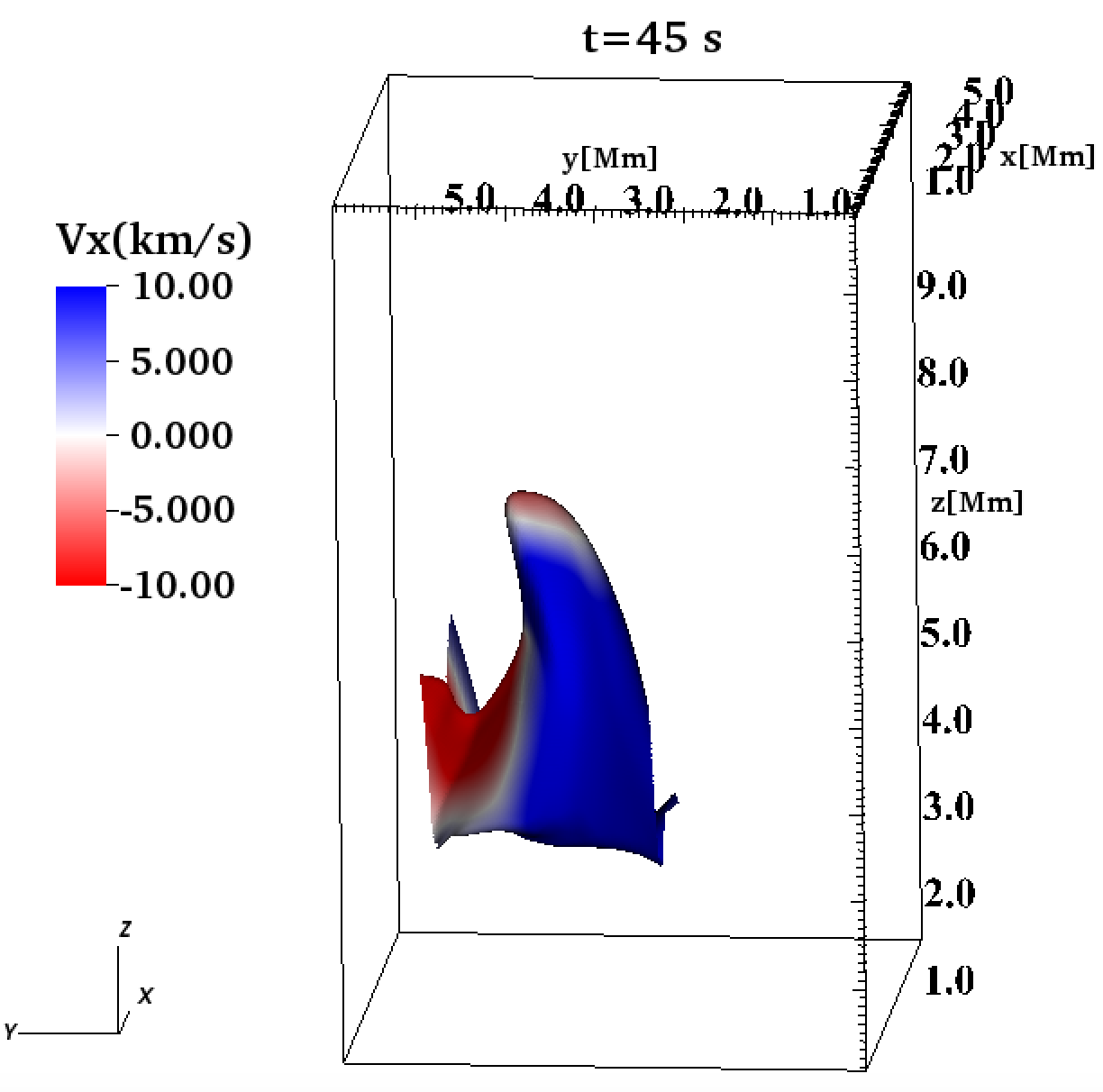}
\includegraphics[width=5.5cm]{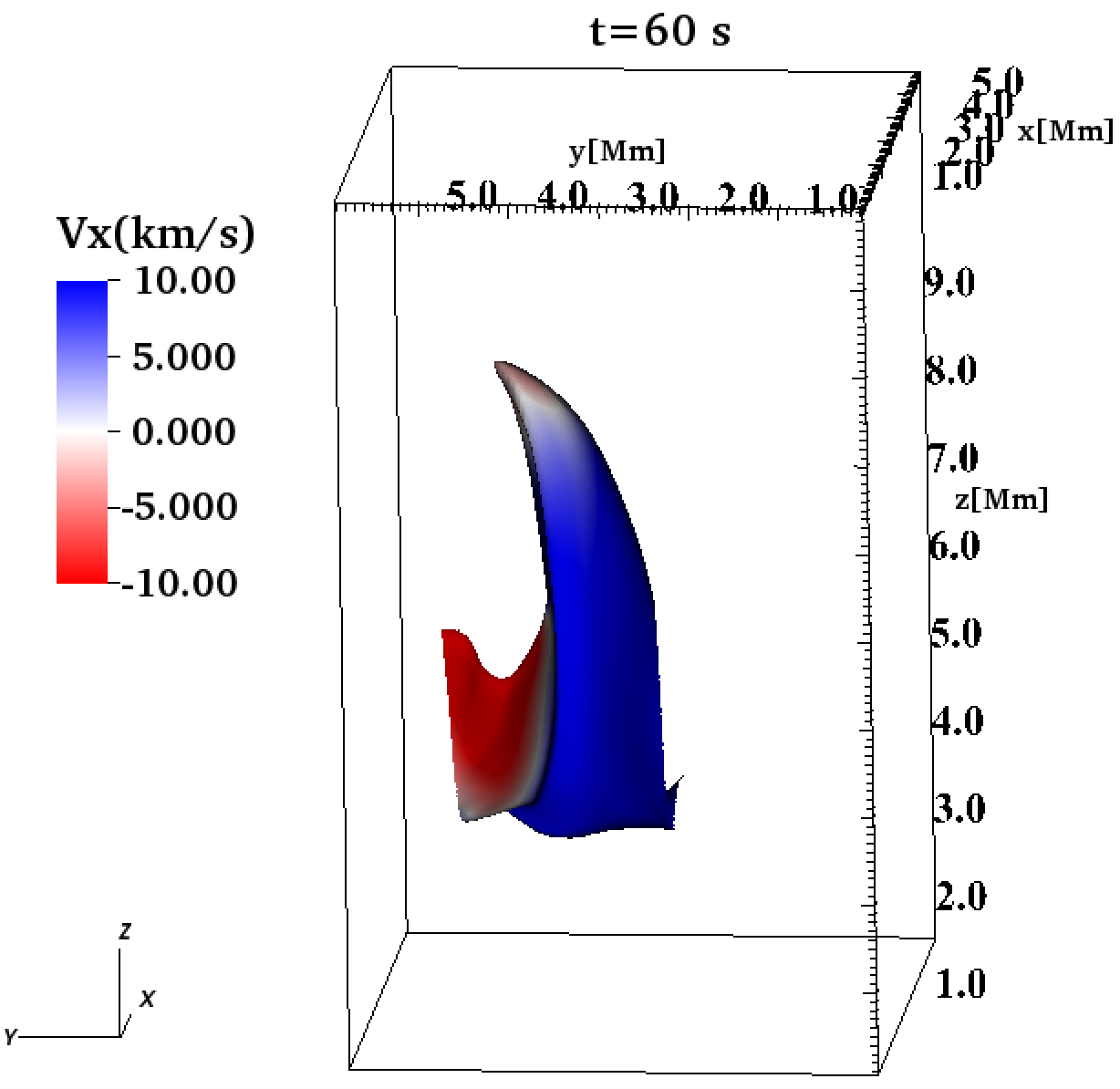}\\
\includegraphics[width=5.5cm]{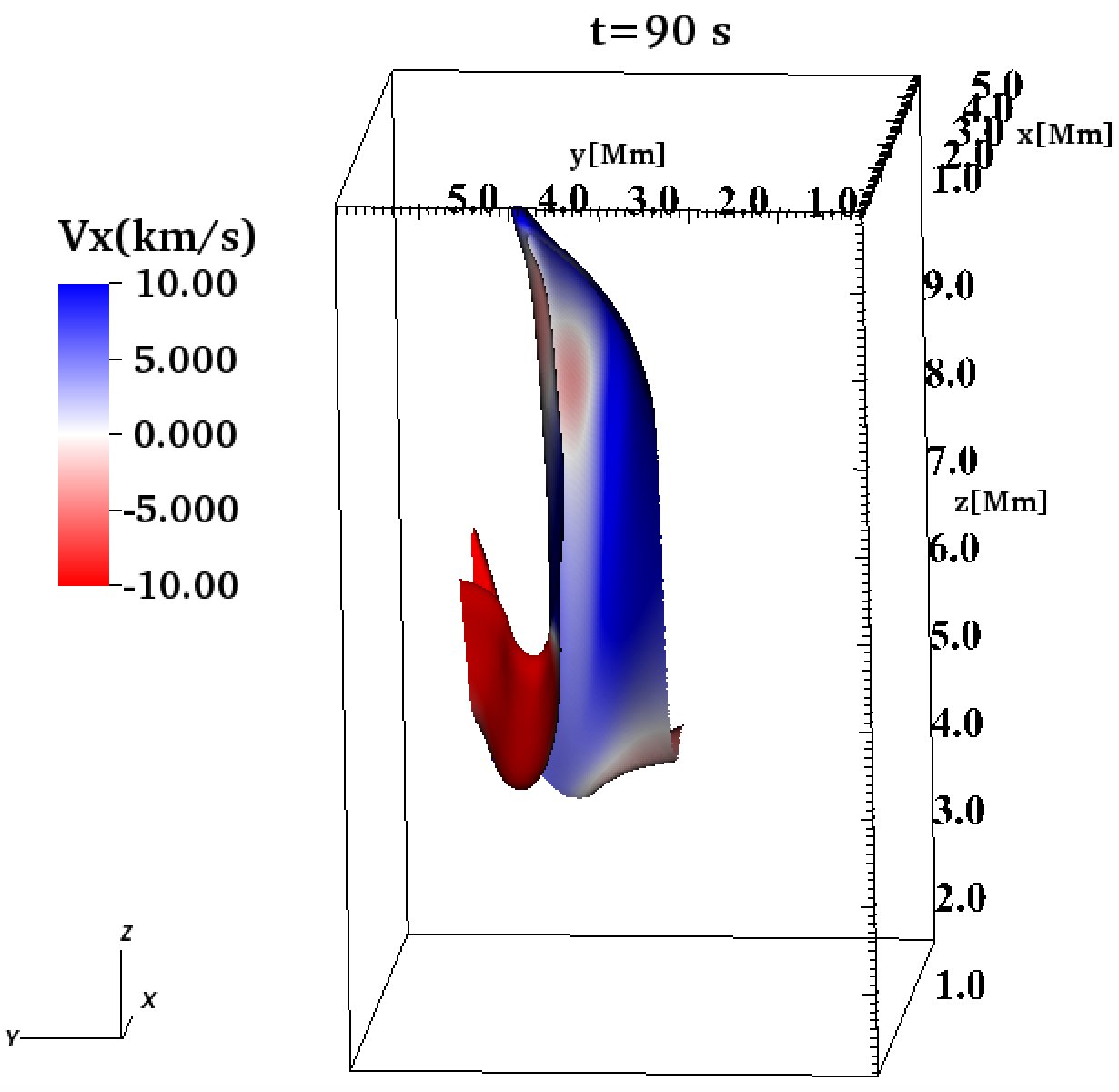}
\includegraphics[width=5.5cm]{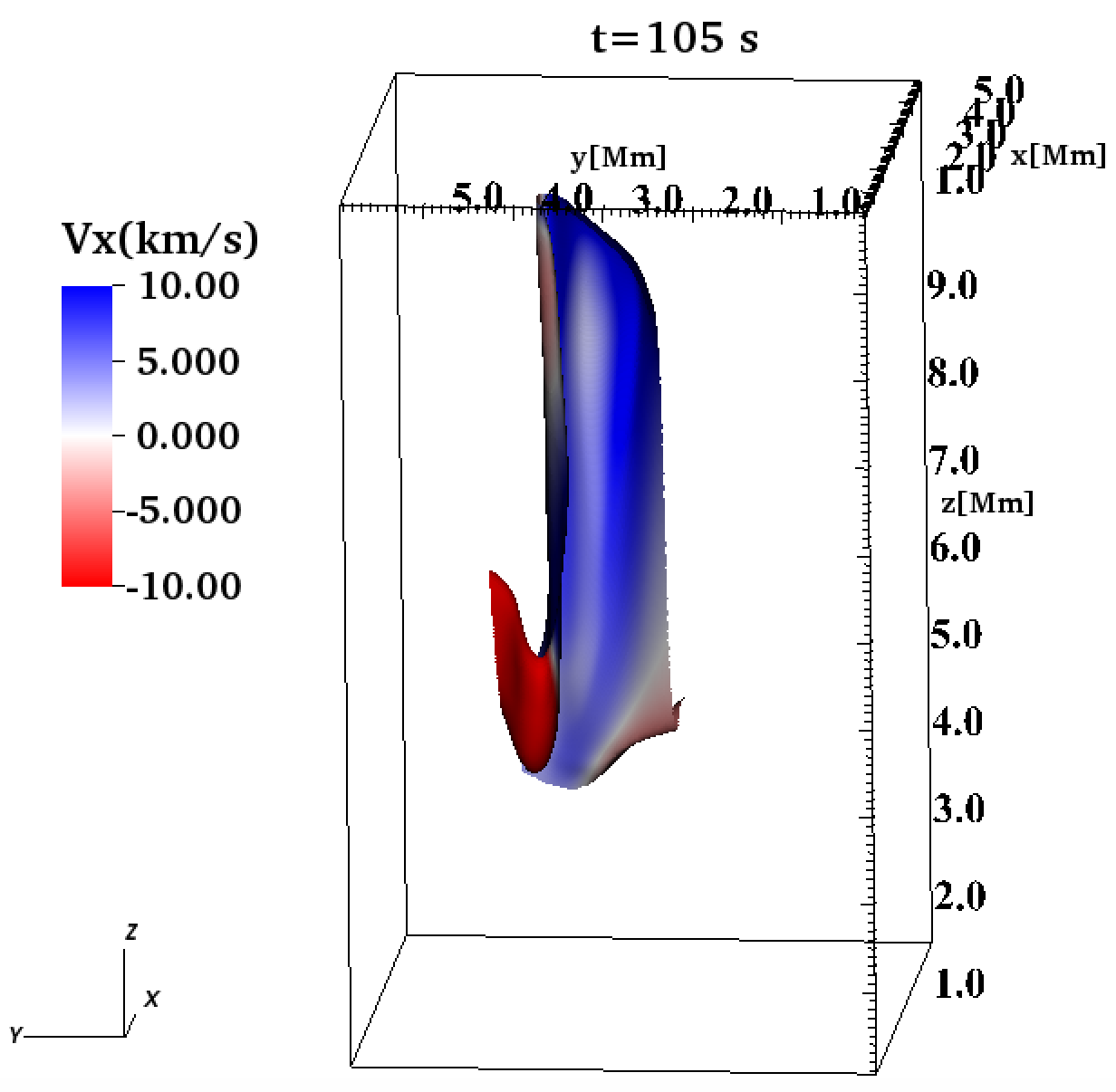}
\includegraphics[width=5.5cm]{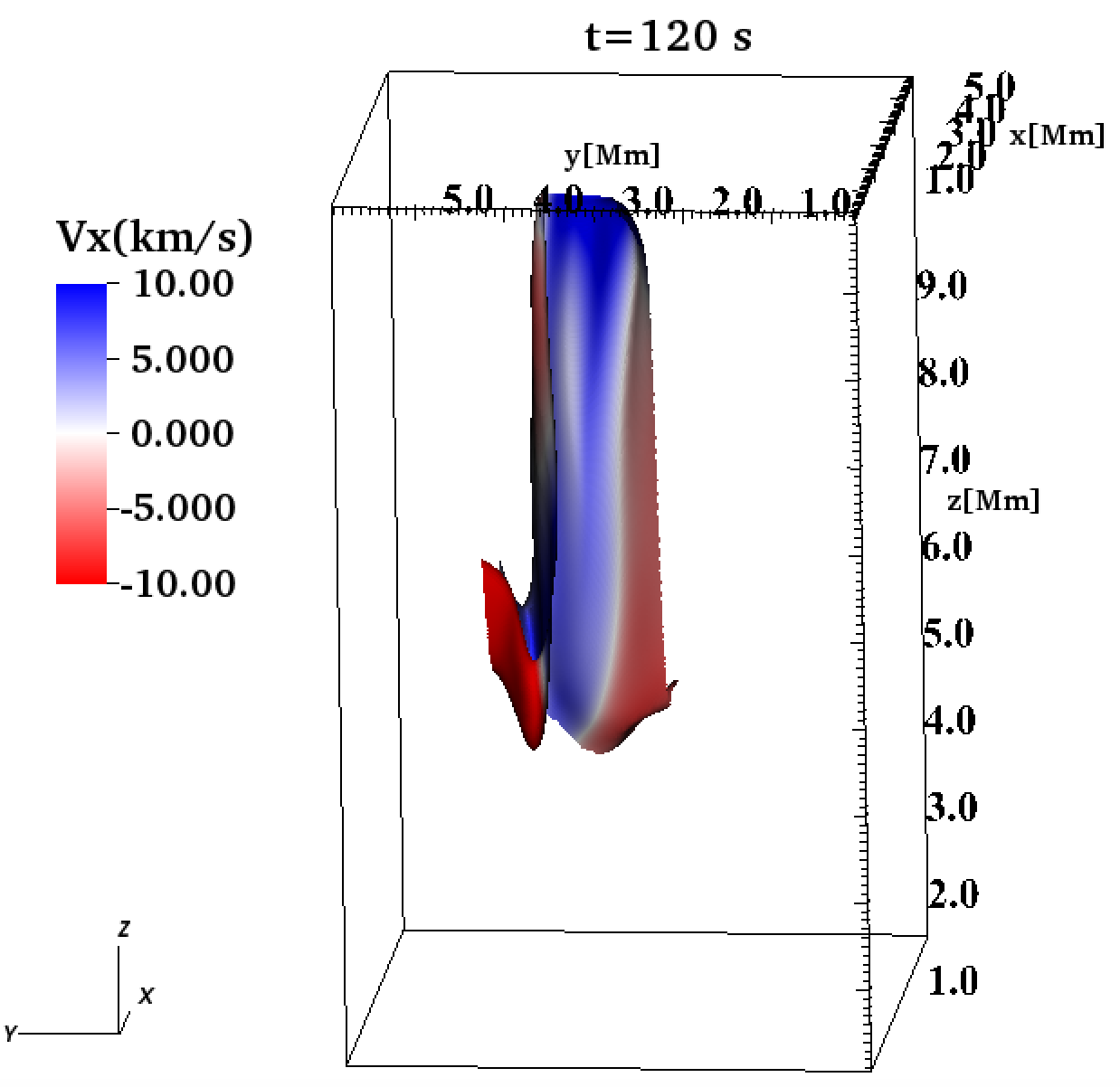}\\
\includegraphics[width=5.5cm]{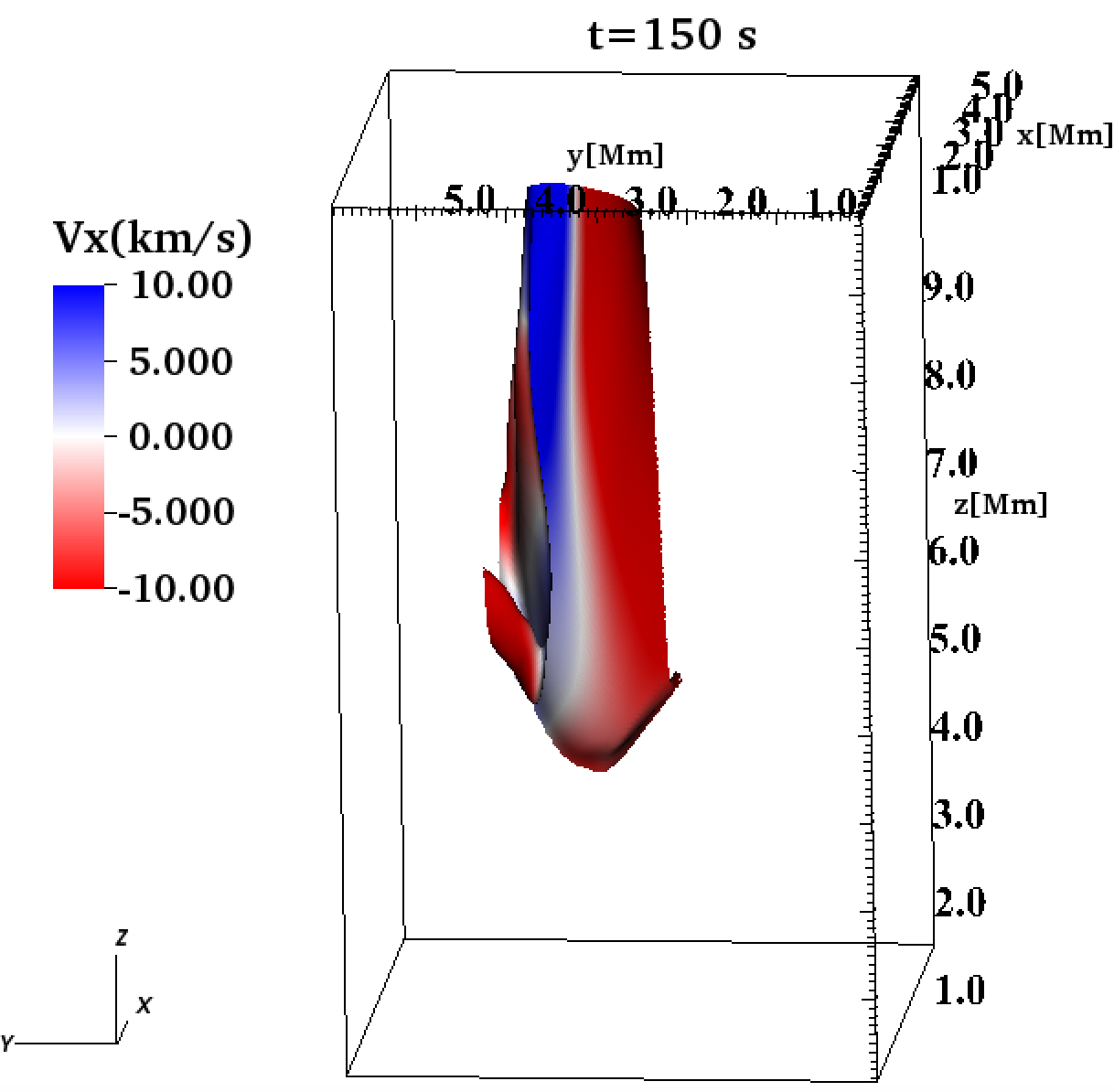}
\includegraphics[width=5.5cm]{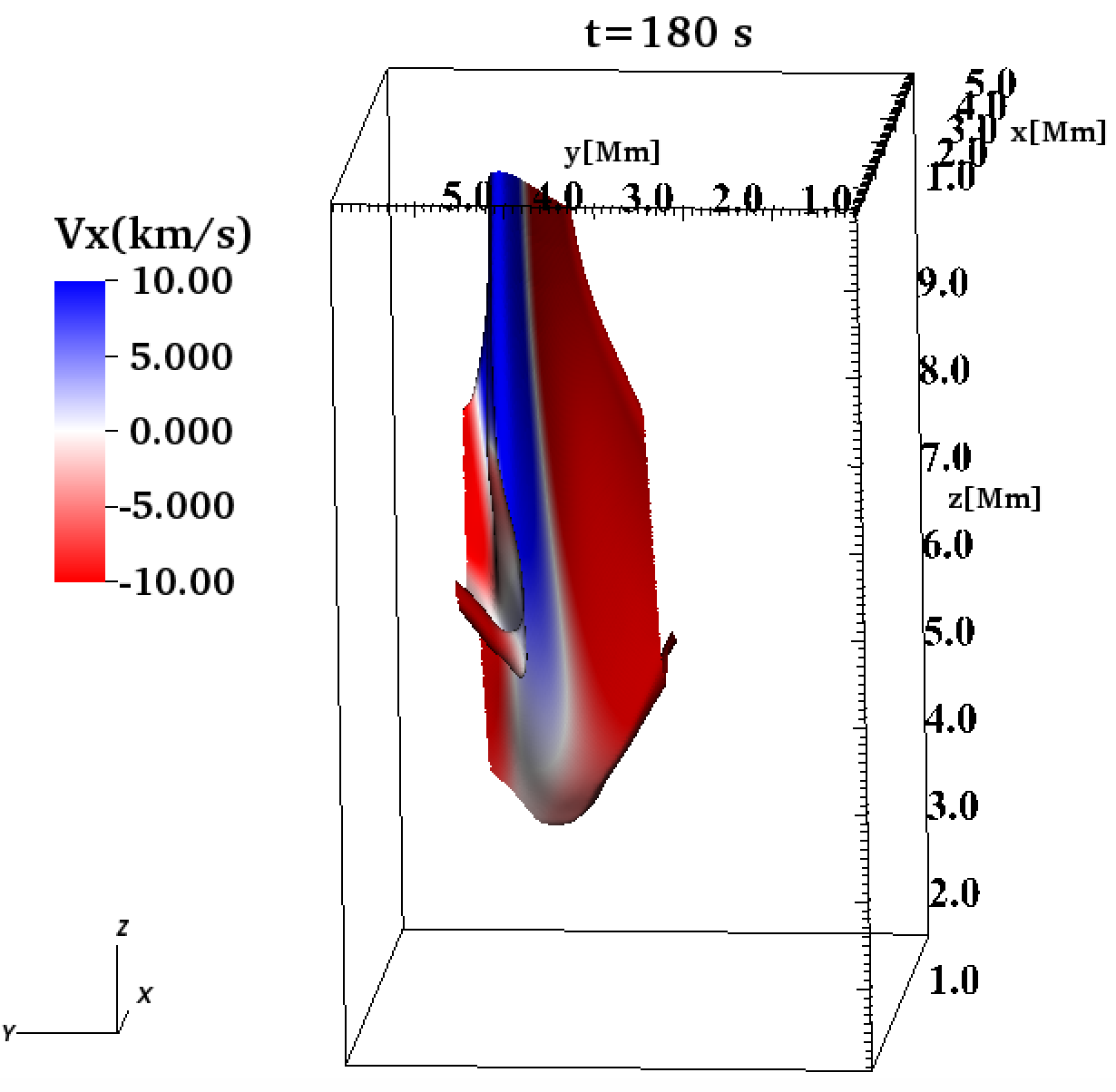}
\includegraphics[width=5.5cm]{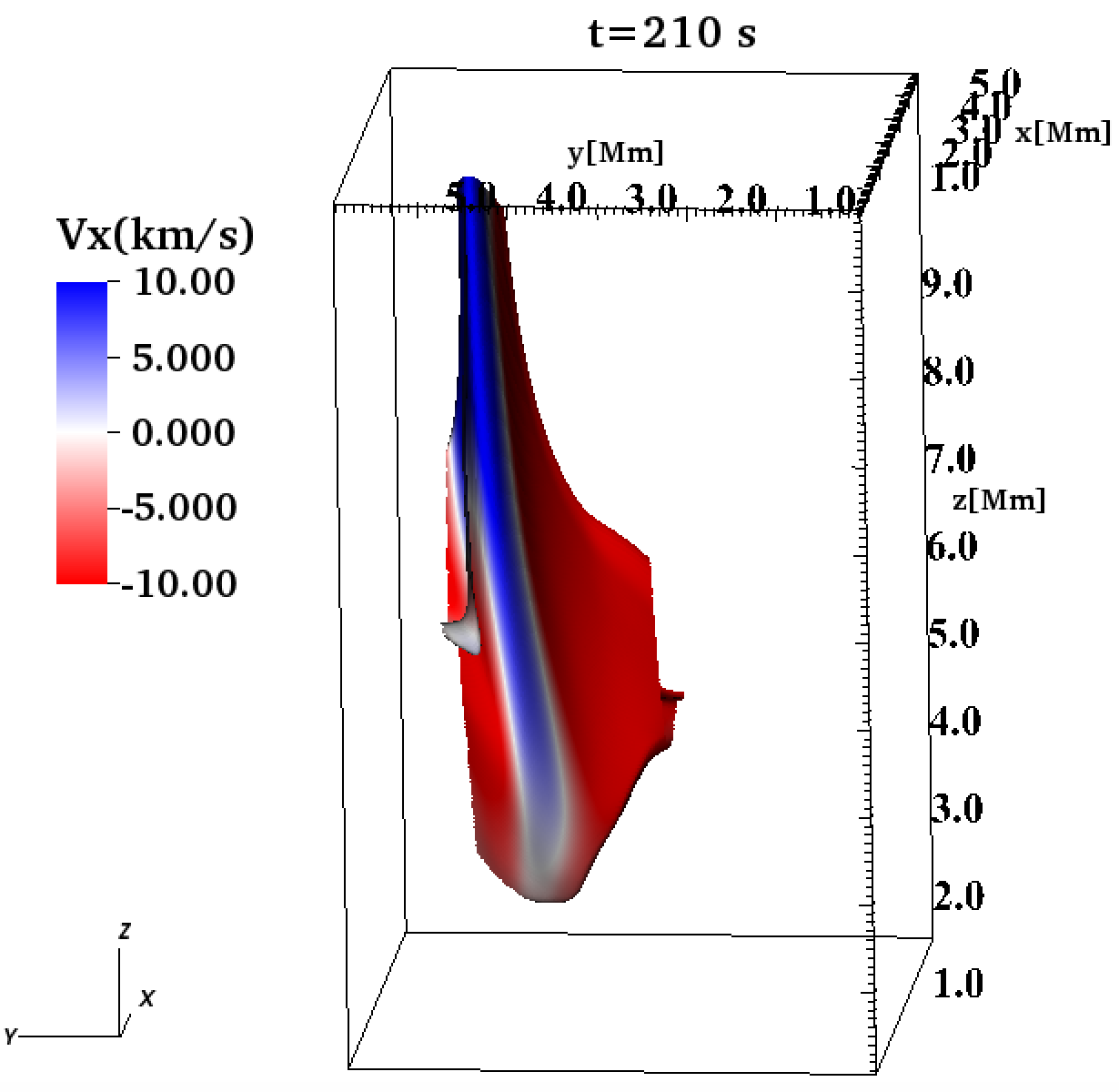}
\caption{\label{3D_temp_contours_vx_color_maps} Snapshots of a temperature contour at various times. The jet is represented  by an isosurface of the plasma temperature equal to $10^{4}$ K. The color
code labels the value of $v_x$. In this perspective  blue indicates motion toward the reader and red toward inside the page.}
\end{figure*}

Another important property of Type II spicules to look at, is whether they are twisted, rotate or show an azimuthal flow component. Doppler shift observations of various emission lines in the limb suggest
that Type II spicules are rotating \citep{De_Pontieu_et_al_2012,Sekse2013,Sharma_2017}. From our simulation it is possible to study the behavior of the velocity components $v_x$ and $v_z$ inside the
jet in order to track possible rotational or twisting motions. A similar analysis was carried out by \citet{Pariat_et_al_2016} to identify torsional/twisting motions of coronal jets. In our case we show
temperature contours with constant value of $10^{4}$ K colored with the distribution of $v_x$ at times $t=30$, 45, 60, 90, 105, 120, 150, 180 and 210 s in Fig. \ref{3D_temp_contours_vx_color_maps}.

\begin{figure*}
\centering
\includegraphics[width=14.0cm]{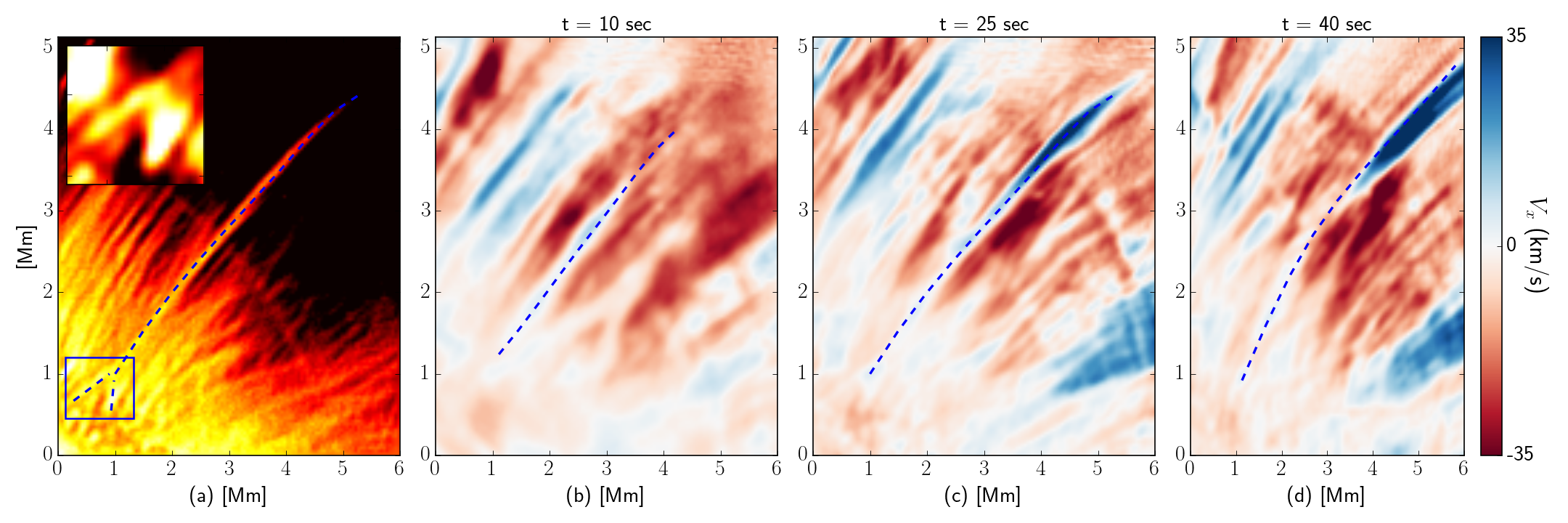}
\caption{\label{doppler_spicule}Left to right: Panels show a spicule (traced as dashed-line) off-limb, observed in H$\alpha$ wavelength (a), with temporal evolution of the line-of-sight (LOS) Doppler velocity
estimates (b-d). The unsharp-masked intensity image (a) show inverted Y-shaped structure (zoomed in inset) at the spicule footpoint (highlighted in box) suggestive of a magnetic reconnection process.
Doppler estimates reveal the longitudinal rise of the spicule with its dominant motion towards the observer (b-c). The development of rotational motion is indicated by the enhanced red-blue asymmetric
profile at the apex of spicule (d).}
\end{figure*}

For the perspective used in this case, the blue color represents motion toward the reader and red color represents motion away from the observer. For instance, at time $t=30$ s the jet starts to develop
and shows both red and blue-shifted plasma. By times $t=45$ and 60 s, the motions are predominantly towards the observer with counter-motion developing at the top of the jet. At time $t=90$ s, the
predominant motion towards the observer and some counter-motion still persists at the top of the jet. This dual behavior lasts through times $t=105$ and $t=120$ s. At times $t=150$, 180 and 210 s the jet
shows a velocity structure represented by a red-blue asymmetry across its width. The time evolution of the jet from the simulation (Fig. \ref{3D_temp_contours_vx_color_maps}) also shows strong
resemblance to the observations of a spicule seen off-limb in H$\alpha$ (Fig. \ref{doppler_spicule}). Details of this observational data, e.g., time cadence and spatial resolution, have been discussed 
previously by \citet{Shetye_et_al_2016}. The unsharp mask intensity image (Fig. \ref{doppler_spicule}(a)) of this spicule suggests it is launched from an inverted Y-shape structure
\citep{Shibata_et_al_2007,He_et_al_2009}, associated with reconnection. The estimated Doppler shift profile ($V_{x}$), at discrete time-steps of the spicule evolution show striking similarities with the
simulated jet. (Fig. \ref{doppler_spicule}(b)) showcases the early rise-phase of the spicule ($t=10$ s) as it starts to penetrate through the ambient chromospheric environment, as seen in Fig.
\ref{3D_temp_contours_vx_color_maps} ($t=30$ s). At the middle-phase of it's evolution (Fig. \ref{3D_temp_contours_vx_color_maps}, $t=105$ s), the spicule attains a mainly blue-shift Doppler profile,
indicating bulk motion towards the observer as shown (Fig. \ref{doppler_spicule}(c)). However, at the late-phase of the spicule's ascent, the apex has developed an asymmetric red-blue Doppler profile (Fig.
\ref{doppler_spicule}(d)), indicating rotational motion, similar to the simulated jet (Fig. \ref{3D_temp_contours_vx_color_maps}, t = 180 s). The rotational motion is prevalent at height above 3 Mm, as is also 
seen in the simulation.

These results together with the time series of the velocity components at points A, B, and C in Fig. \ref{velocity_components_over_time}, clearly indicate rotational motions of the jet.

\subsection{Vertical motions} 

\begin{figure*}
\centering
\includegraphics[width=5.6cm]{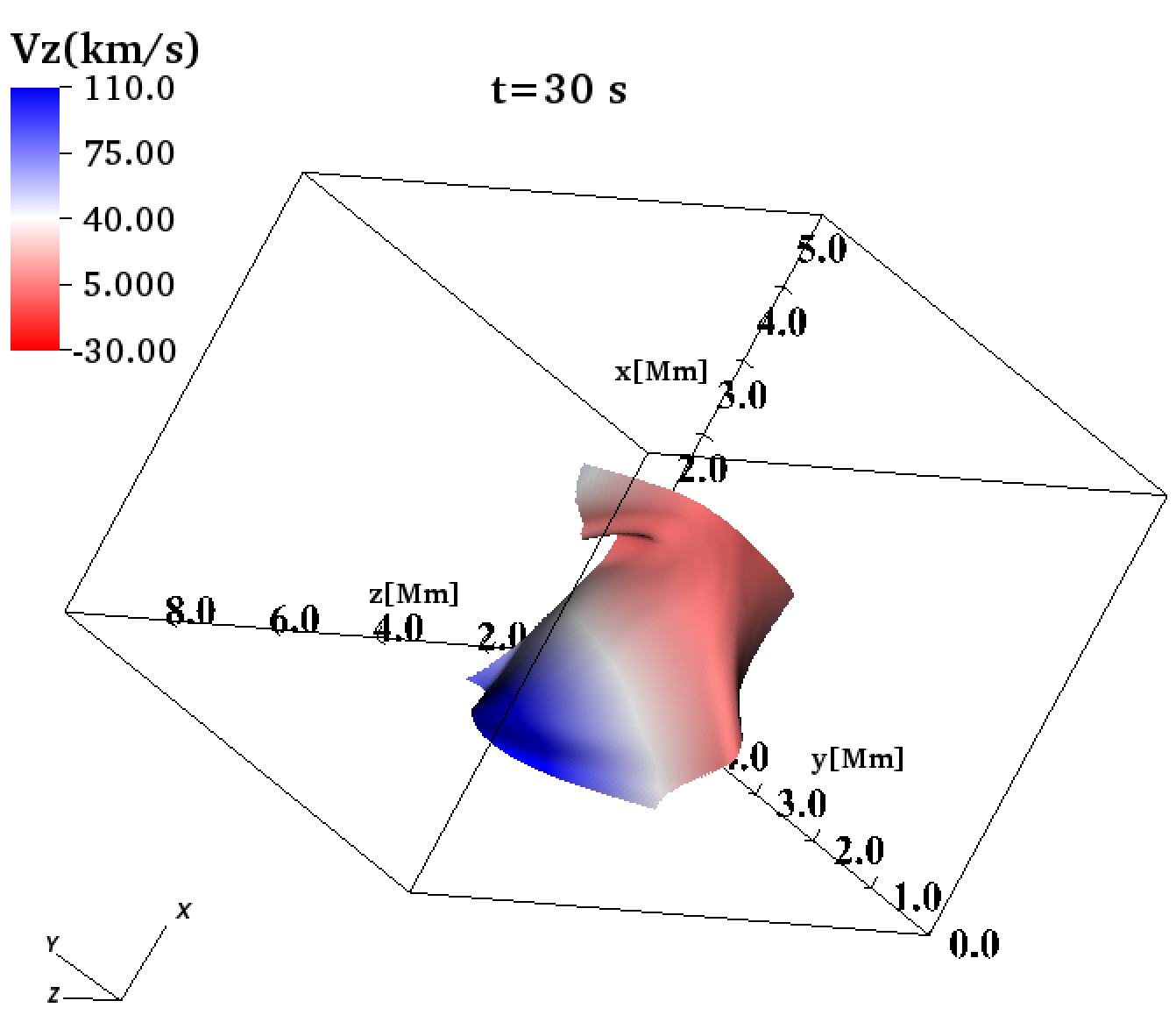}
\includegraphics[width=5.6cm]{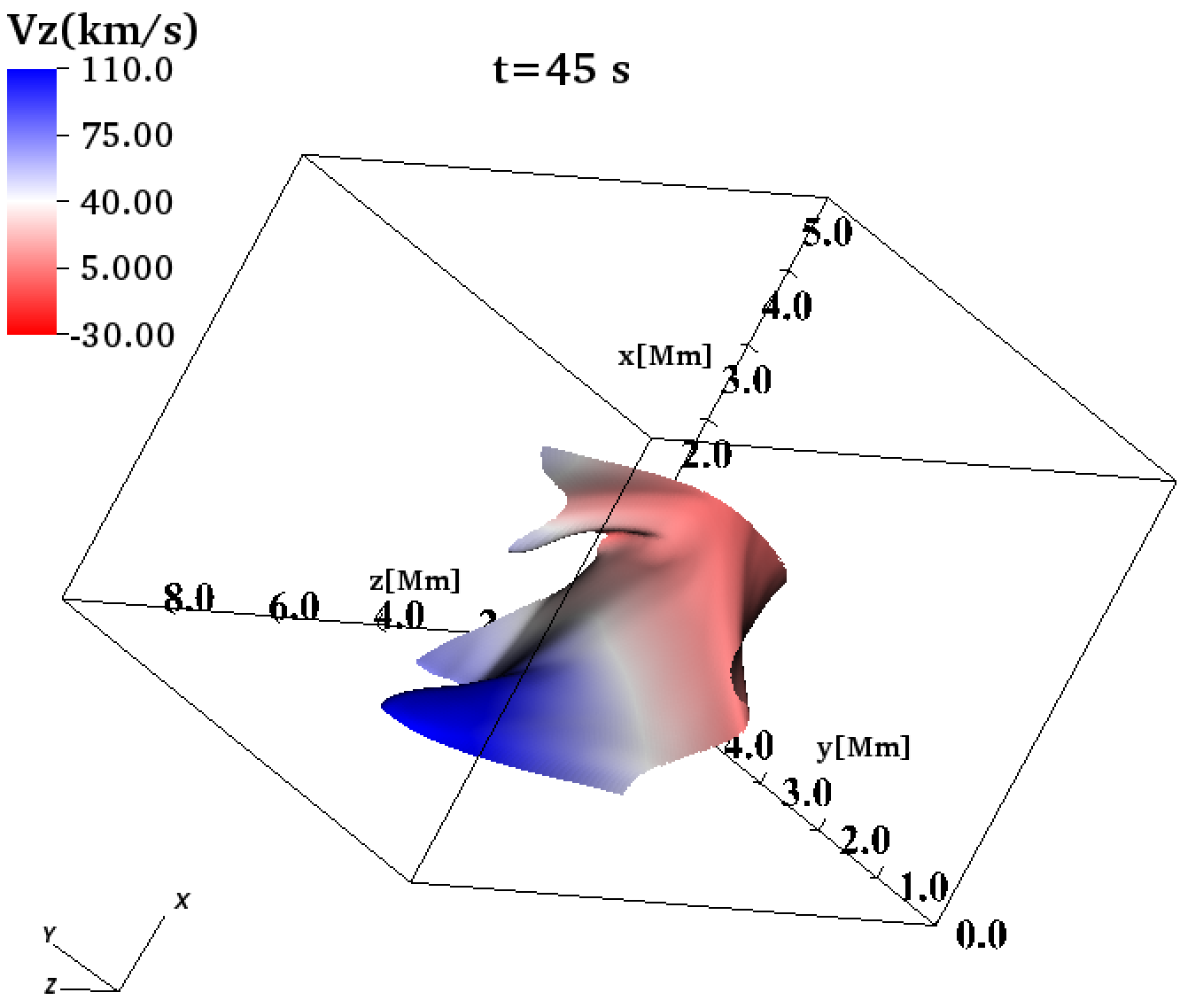}
\includegraphics[width=5.6cm]{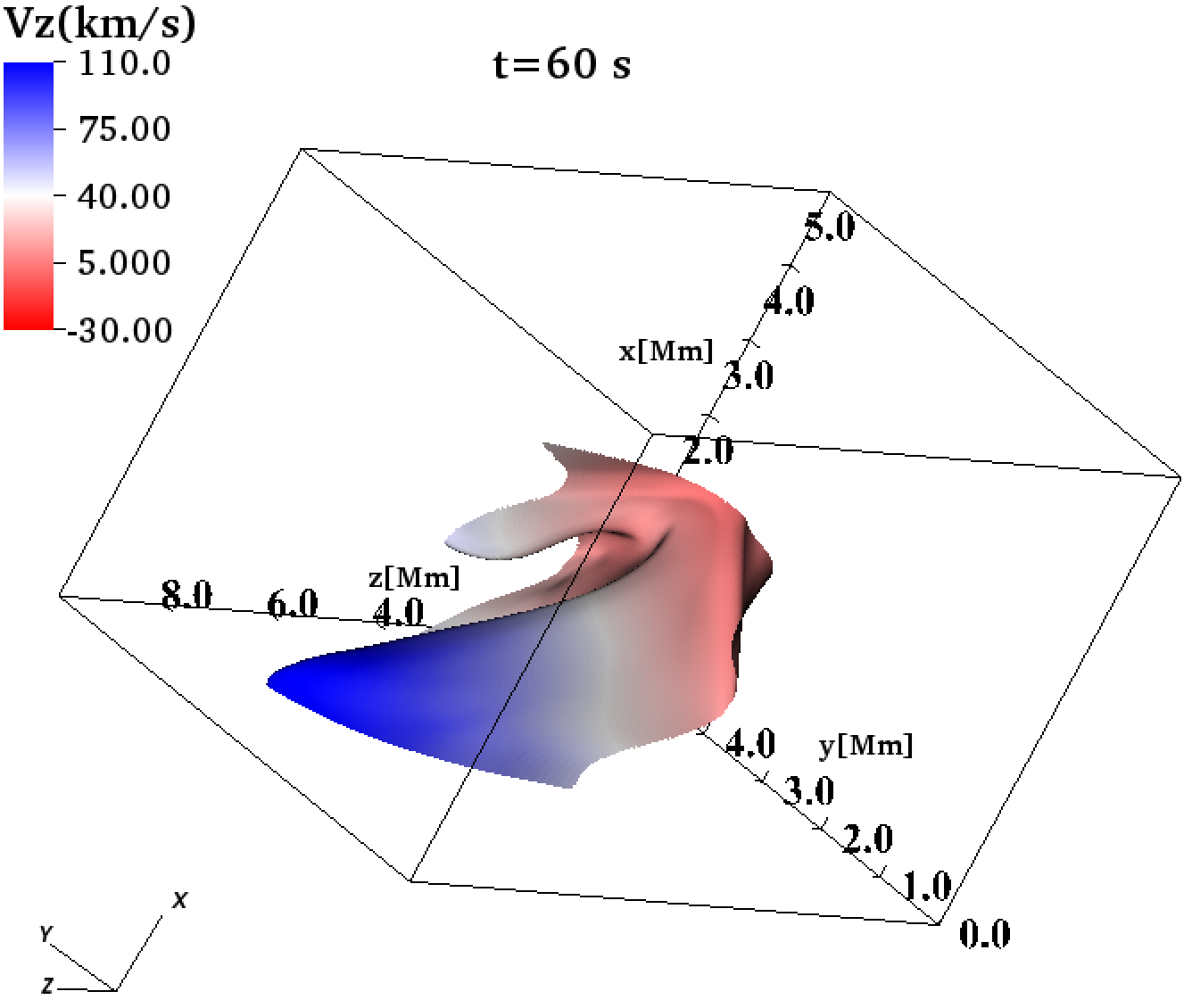}\\
\includegraphics[width=5.6cm]{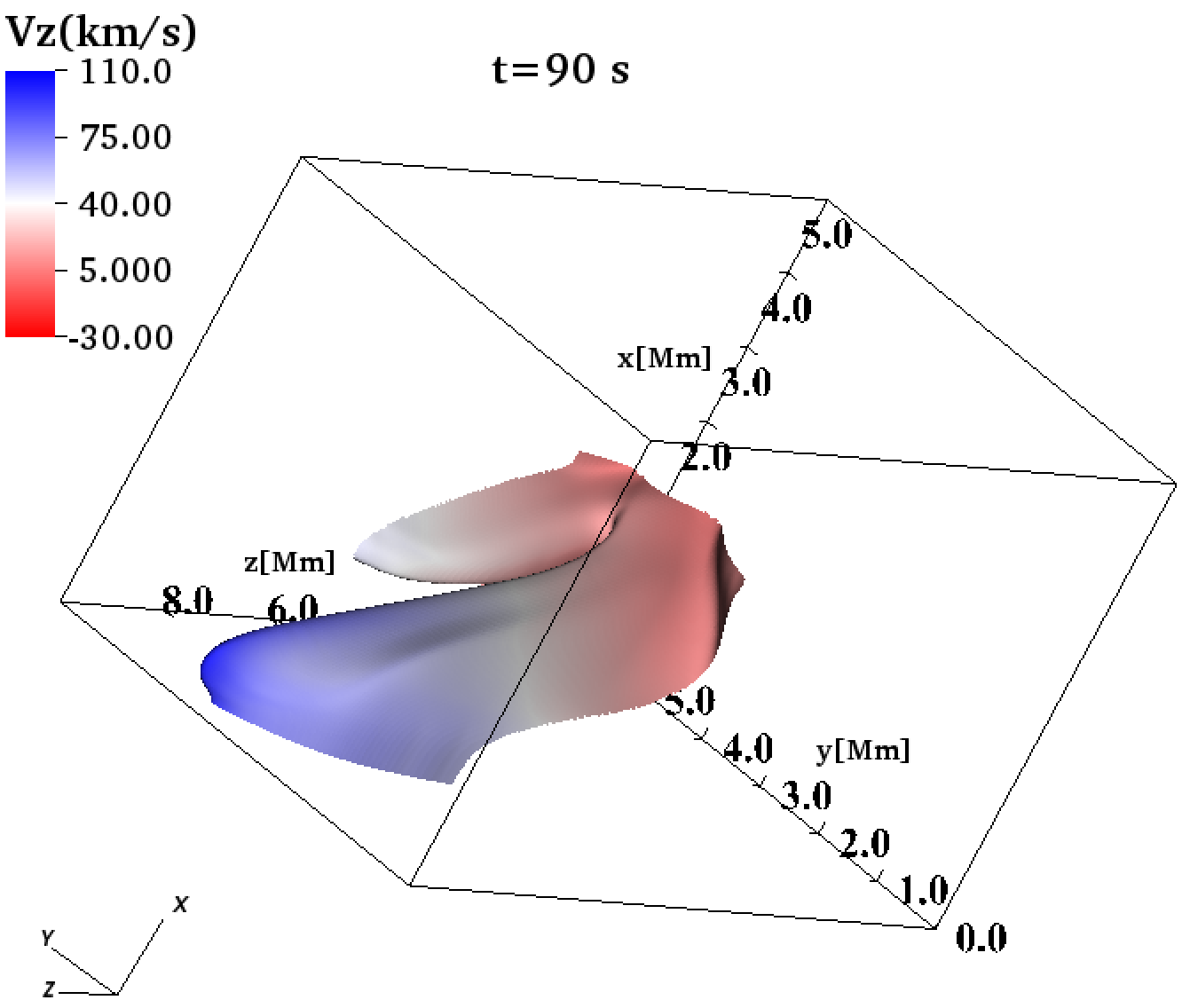}
\includegraphics[width=5.6cm]{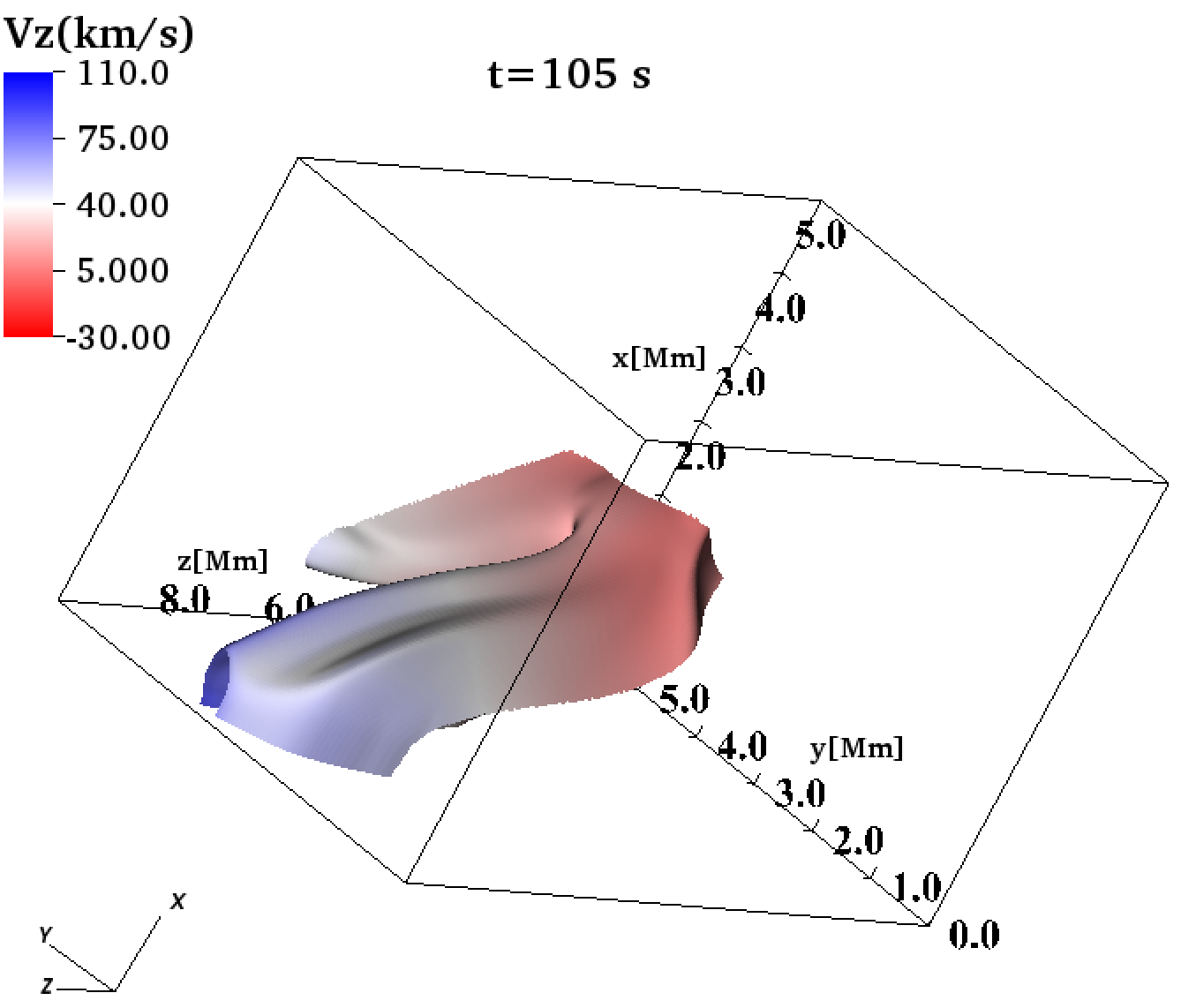}
\includegraphics[width=5.6cm]{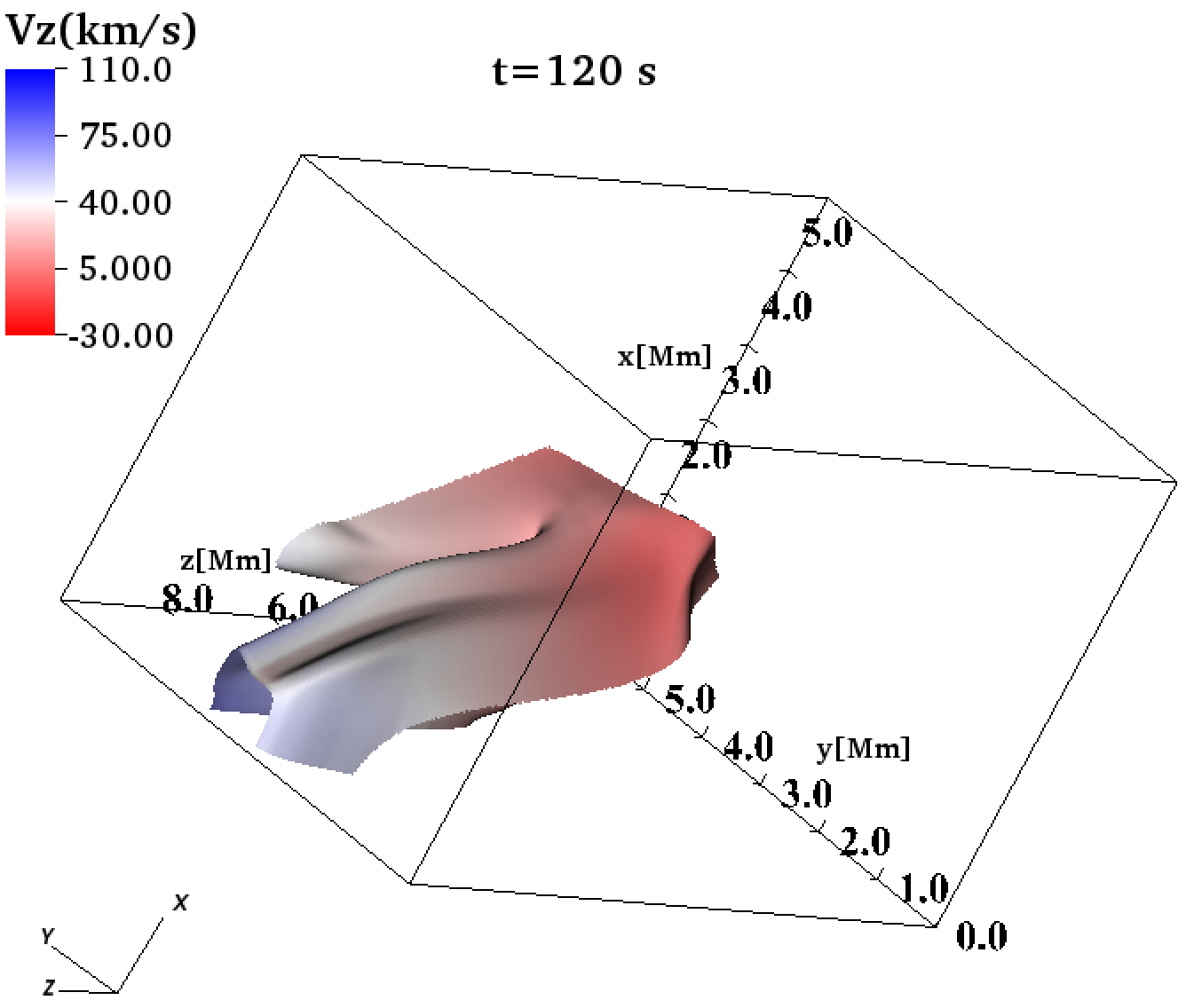}\\
\includegraphics[width=5.6cm]{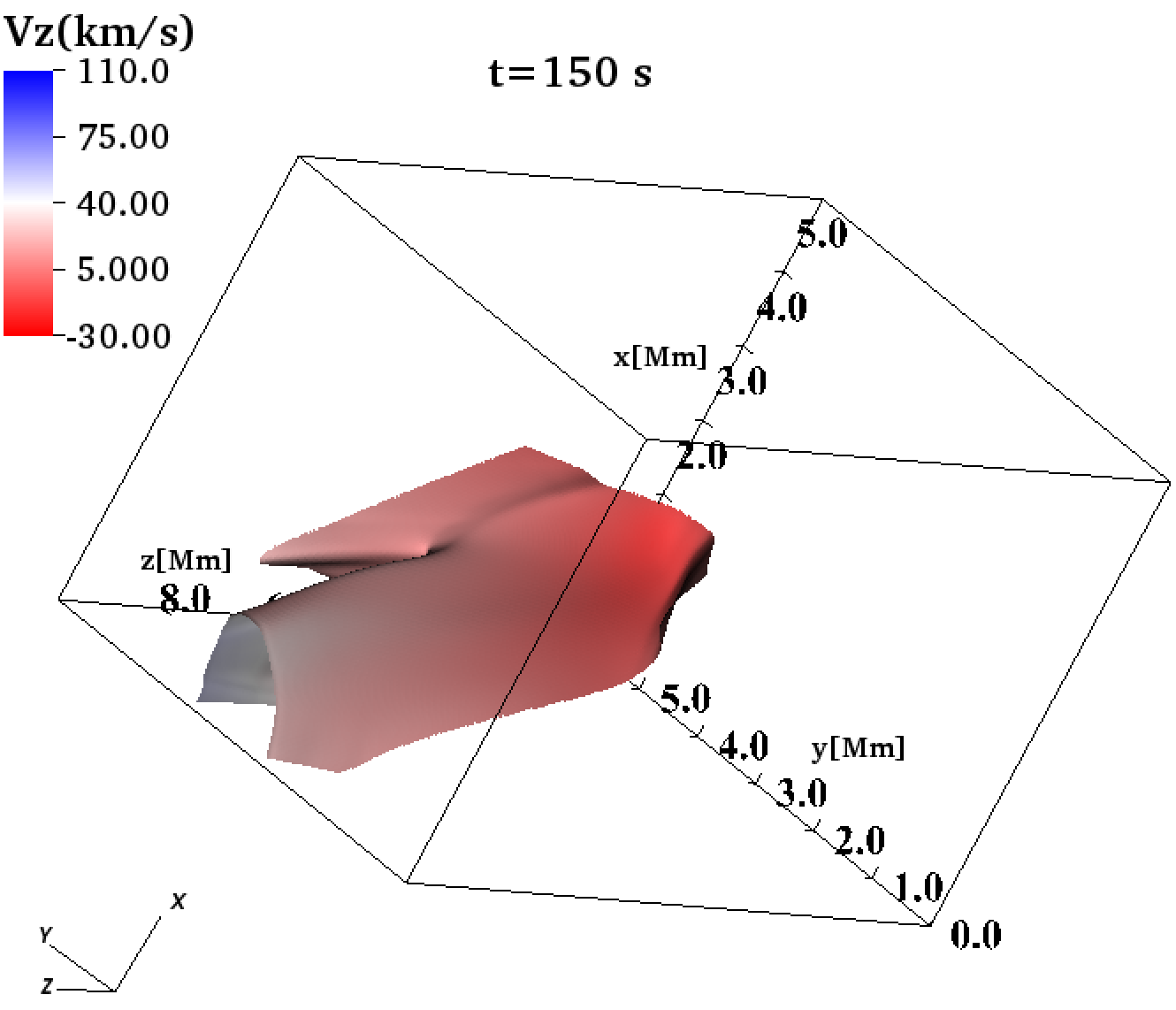}
\includegraphics[width=5.6cm]{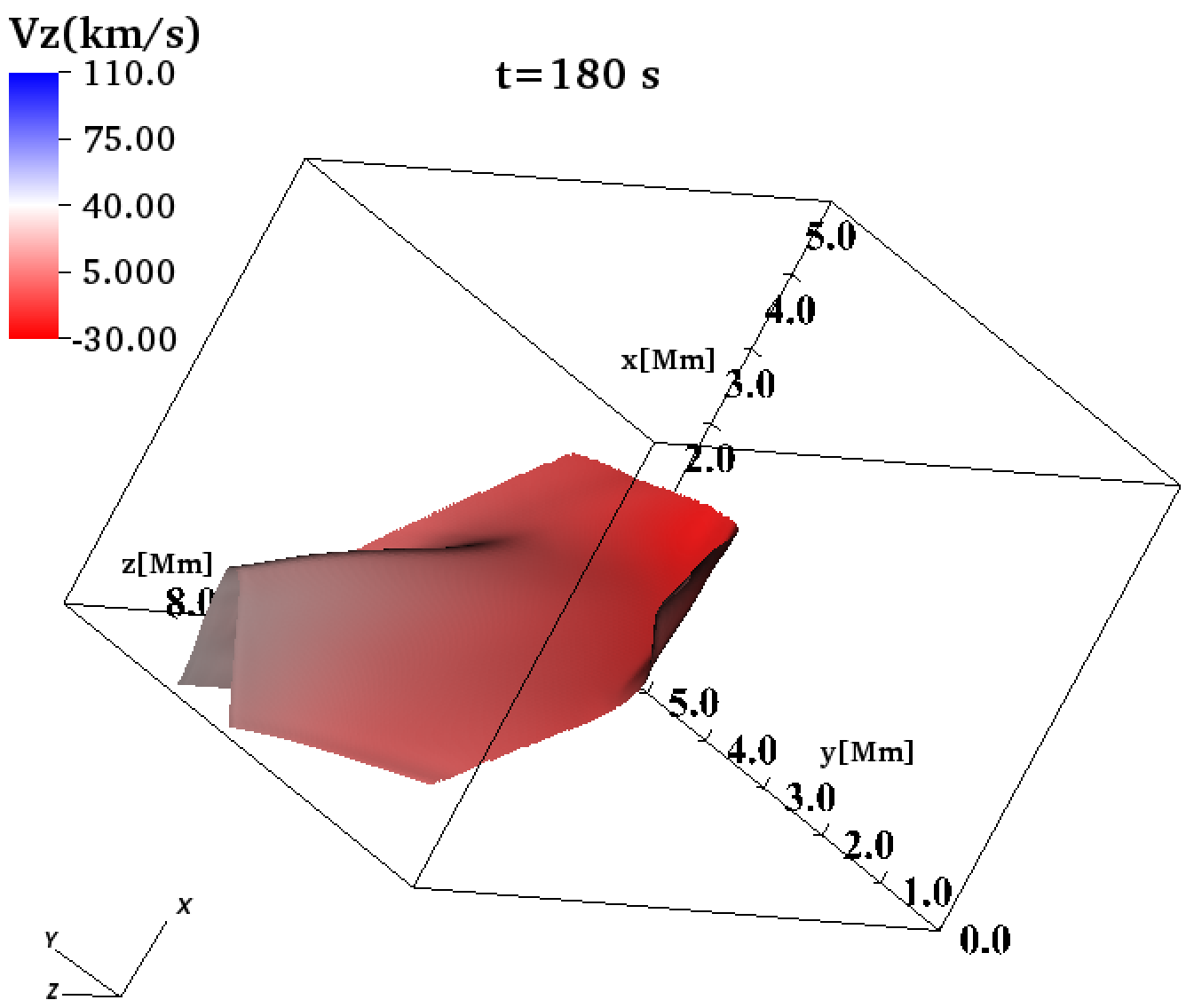}
\includegraphics[width=5.6cm]{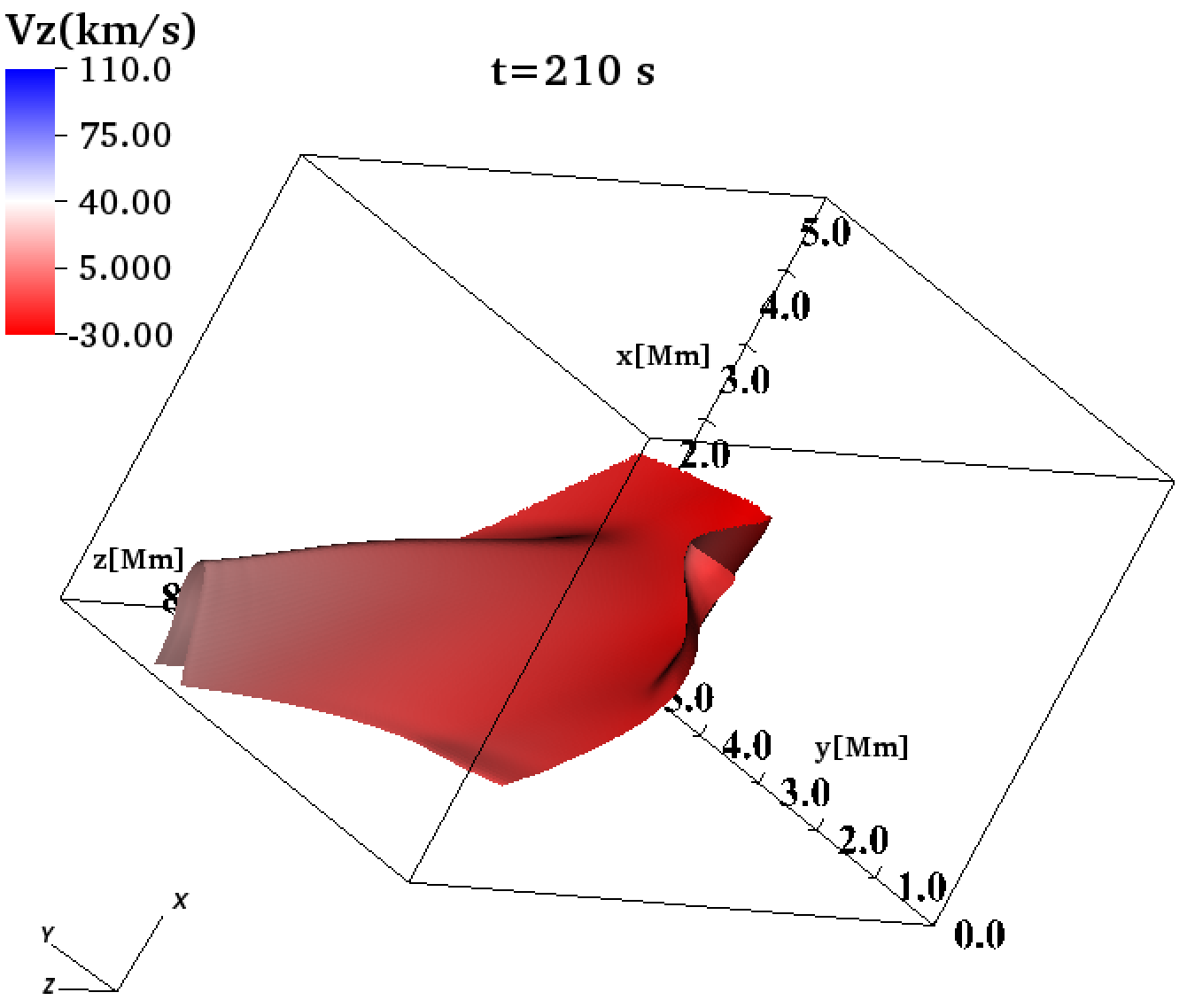}
\caption{\label{3D_temp_contours_vz_color_maps} 
Snapshots of a temperature contour at various times. The jet is represented  by an isosurface of the plasma temperature equal to $10^{4}$ K. The color-code labels the value of $v_z$.  The color red and
blue indicates downward and upward velocity respectively.}
\end{figure*}

In Fig. \ref{3D_temp_contours_vz_color_maps} we show the isosurface of temperature colored with the values of $v_z$, which helps to track the vertical motion of the jet, for instance at times $t=30$, 45
and 60 s, the jet practically shows upward motion from the middle to the top. By times $t=90$, 105 and 120 s, the amplitude of vertical motion start to decrease at the bottom of the jet. Finally, at times
$t=$150, 180 and 210 s the jet starts moving downwards, in particular this behavior is consistent with the observed vertical motion with velocities of  order $110$ km s$^{-1}$ in Type II spicules
\citep{Skogsrud_et_al_2014}. 

\begin{figure}
\centering
\includegraphics[width=5.0cm]{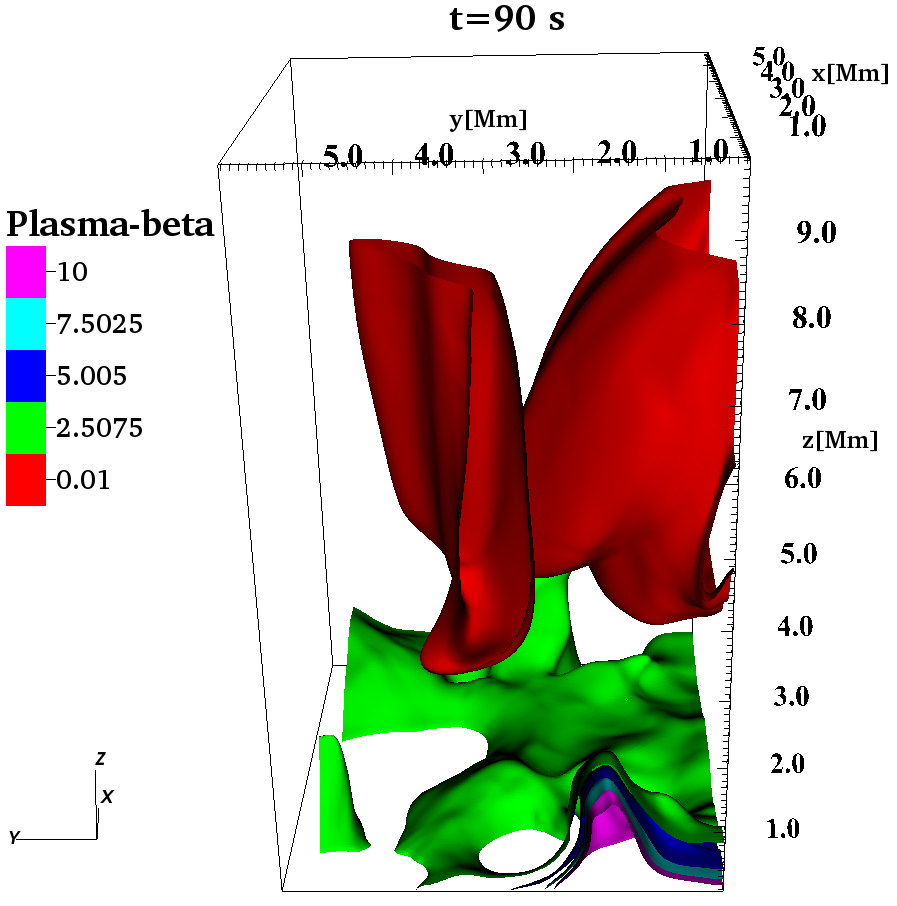}
\caption{\label{Plasma-beta_3D_contours} Contours of plasma $\beta$ at time t=90 s. The contours indicate that magnetic field is dominant in the region where the spicule is formed as is shown in Fig.
\ref{3D_temp_contours_vx_color_maps}.}
\end{figure}


\section{Conclusions}
\label{sec:conclusions}

In this paper by analyzing temperature isosurfaces to localize the jet, together with the analysis of the horizontal velocity components, we find that the development of a red-blue asymmetry across the jet is
due to rotational motion. Interestingly, the rotational motion is initially clockwise and then begins to move in an anti-clockwise direction, indicating the presence of torsional motion. By analyzing the time
series of $v_x$ and $v_y$ at points near and within the jet at different heights we showed that the rotational motion is generated in its upper region. In addition, by calculating a time-distance plot of the
logarithm of temperature in a horizontal cross-cut at a height of 7 Mm it was shown that the jet also undergoes a considerable transverse displacement.  
 
Additionally, we have presented observational support of rotational motion in an off-limb spicule appearing in the corona (and not being generated from below) in Fig. \ref{doppler_spicule}(d). We can also
see the simulated jet has a dual behavior (i) transverse motion at the foot (0-3 Mm) and (ii) twisted motion at the middle and top parts (3-10 Mm). The rotational type motion (initially clockwise and after anti
clockwise) can be interpreted as  torsional starting at the top of the jet, when it reaches a region where the magnetic field dominates $\beta < 1$ as shown in Fig. \ref{Plasma-beta_3D_contours} and the
Lorentz force is also bigger than pressure gradients $|{\bf J}\times{\bf B}|>|\nabla{p}|$ as shown in Fig. 7 of Paper \citet{Gonzalez-Aviles_et_al_2018}. This is important as it shows that torsional waves can
be generated directly in the  corona and therefore the whole wave energy (i.e without any loses due to propagation from the photosphere and dynamic chromosphere to the corona, as is usually suggested can be dissipated in the corona. For example, regions with ($\beta < 1$) are perfect for the decay of torsional Alfv\'en waves into kinetic Alfv\'en waves, see e.g. cross-scale nonlinear coupling and plasma energization by Alfv\'en waves \citep{Voitenko&Goossens_2005}, excitation of kinetic Alfvén turbulence by MHD waves and energization of space plasmas \citep{Voitenko&Goossens_2004} or the transformation of MHD Alfvén waves in space plasma \citep{Fedun_et_al_2004}.
 
From a nearly vertical perspective of the jet, the vertical component of the velocity shows a blue-red shift, that is similar to the observed in the transition region and coronal lines as shown in Fig. 18 of
\citet{Martinez-Sykora_et_al_2013}, where the Doppler shifts correspond to velocities within the range -8 to 8 km s$^{-1}$. Finally, although there is no magnetoconvection, in the simulated plasma jet we
conclude that rotational motion can still occur naturally at coronal heights without the need of  any photospheric driver, e.g., granular buffeting or vortex motion. In fact, we have shown that such jets could
be an in-situ driver of torsional Alfv\'en waves in the corona.


\section*{Acknowledgments}

This research is partly supported by the following grants: Royal Society-Newton Mobility Grant NI160149, CIC-UMSNH 4.9, and CONACyT 258726 (Fondo Sectorial de Investigaci\'on para la Educaci\'on).
The simulations were carried out in the Big Mamma cluster at the LIASC-IFM. V.F. and G.V. thank the STFC for their financial support. J.-G gratefully acknowledges DGAPA postdoctoral grant to
Universidad Nacional Aut\'onoma de M\'exico (UNAM). Visualization and analysis of the simulations data was done with the use of the VisIt software package. The authors thank to G. Doyle and E. Scullion
for providing the observational data which was collected using the Swedish 1-m Solar Telescope. This Telescope is operated on the island of La Palma by the Institute for Solar Physics of Stockholm University in the Spanish Observatorio del Roque de los Muchachos of the Instituto de Astrofísica de Canarias. The Institute for Solar Physics is supported by a grant for research infrastructures of national importance from the Swedish Research Council (registration number 2017-00625). The authors also wish to acknowledge the DJEI/DES/SFI/HEA Irish Centre for High-End
Computing (ICHEC) for the provision of computing facilities and support. This work also greatly benefited from the discussions at the ISSI workshop "Towards Dynamic Solar Atmospheric Magneto
Seismology with New Generation Instrumentation".






\bsp	
\label{lastpage}
\end{document}